\documentclass[fleqn,usenatbib]{mnras}

\usepackage{newtxtext,newtxmath}

\usepackage[T1]{fontenc}
\usepackage{soul}

\DeclareRobustCommand{\VAN}[3]{#2}
\let\VANthebibliography\thebibliography
\def\thebibliography{\DeclareRobustCommand{\VAN}[3]{##3}\VANthebibliography}

\usepackage{graphicx}	
\usepackage{amsmath}	

\usepackage{xcolor}

\title[Fourier Analysis of Galactic Bars]{Characterization of Orbits in Bars in Disc Galaxies Using Fourier Frequencies}

\author[A. Silva-Castro,  I. Puerari and D. Valencia-Enríquez]{
A. Silva-Castro$^{1}$\thanks{E-mail: alan.silvacastro98@gmail.com},
I. Puerari$^{1}$ and
D. Valencia-Enríquez$^{2}$
\\
$^{1}$Instituto Nacional de Astrofísica, Óptica y Electrónica, Calle Luis Enrique Erro, No. 1, 72840, Santa María Tonantzintla, Puebla, México.\\
$^{2}$Universidad Mariana, Calle 18 No. 34-104, 52001 Pasto, Colombia\\
}

\date{Accepted September 8, 2026. Received September 4, 2026; in original form November 11, 2025}

\pubyear{\the\year{}}

\begin{document}
\label{firstpage}
\pagerange{\pageref{firstpage}--\pageref{lastpage}}
\maketitle

\begin{abstract}

Bars are among the most prominent dynamical structures in disc galaxies, yet their long-term evolution and the orbital families that sustain them remain only partially understood. In this work, we introduce a new methodology based on frozen potentials combined with Fourier analysis of particle trajectories to identify regular and sticky orbits in self-consistent $N$-body simulations. This approach allows us to isolate particles belonging to the $x_1$ family, directly associated with the bar, and the $x_2$ family, linked to a secondary perpendicular structure. Tracking these orbits across multiple snapshots enables us to quantify the time evolution of key bar properties, including its mass fraction, semi-axes, and density profile. Compared with established used diagnostics, our method provides the closest match to the temporal behaviour of the Fourier bar-strength parameter $A_2$, while offering a direct dynamical interpretation and full three-dimensional characterization that Fourier-based approaches cannot provide. Our analysis further reveals that the indices $n_x$, $n_y$, and $n_z$ describing the bar density distribution are anisotropic and rarely equal to the canonical $n=2$ assumed in Ferrers profiles, highlighting the limitations of standard analytical approximations. We also find strong correlations between bar growth, orbital chaoticity, and angular momentum redistribution, with the bar acting both as a sink and as a driver of angular momentum transfer. A comparison between the two analysed models shows that rapidly forming bars are weaker and less stable than those that grow more gradually. Altogether, our methodology provides a robust framework for linking orbital structure to the dynamical evolution of barred galaxies.

\end{abstract}




\section{Introduction}



Understanding the formation and evolution of disc galaxies is one of the central goals of extragalactic astrophysics. Among the internal structures that shape their dynamics and long-term evolution, stellar bars play a prominent role. Galactic bars are elongated stellar structures found in a significant fraction of disc galaxies, both in the local universe and at high redshift \citep{2004ApJ...612..191E, 2012ApJ...757...60K, 2014MNRAS.438.2882M, 2023ApJ...947...80B, 2024MNRAS.530.1984L, 2025MNRAS.542..151M, 2025ApJ...987...74G}. Over the years, observational efforts have greatly improved our understanding of their properties. Studies have characterized the bar length and strength and analysed how these quantities vary across different types of host galaxies \citep{2005MNRAS.364..283E, 2016A&A...587A.160D}. Furthermore, bars are known to influence several aspects of galactic evolution, including star formation activity, stellar populations, color, and metallicity \citep{2016A&A...595A..63V}. Their formation and structural properties have also been examined in the context of the galactic environment, such as whether residing in dense clusters affects their prevalence or morphology \citep{2010ApJ...711L..61M, 2023A&A...679A...5A}.

Although observational data is fundamental, it faces inherent limitations due to the vast timescales involved in galactic evolution, the large distances to galaxies, and the constraints of current technology. For instance, directly measuring dynamical quantities such as the pattern speed $\Omega_p$ and the ratio $\mathcal{R}=R_{\text{CR}}/R_{\text{Bar}}$, where $R_{\text{CR}}$ is the corotation radius and $R_{\text{Bar}}$ is the bar radius, is notoriously challenging and often subject to large uncertainties \citep{2015A&A...576A.102A, 2022MNRAS.517.5660G}. In this context, numerical simulations have become essential tools in the study of galactic dynamics, providing access to physical processes and structures that cannot be observed directly. For example, even basic morphological properties such as bar length can differ between simulations and observations, highlighting the importance of understanding and reconciling both perspectives \citep{2005MNRAS.364..283E, 2024MNRAS.531..751P}. This makes simulations an ideal framework for investigating bar formation, evolution, and the underlying dynamical mechanisms driving their secular development.

The formation of galactic bars can be triggered by various mechanisms, most notably internal disc instabilities. These are often linked to the mass distribution within the galaxy, such as a dominance of baryonic matter over dark matter in the central regions \citep{2019AJ....157..175V, 2023ApJ...947...80B}. Regardless of the initial trigger, bar formation is closely associated with a dynamical process known as apsidal precession synchronization, in which stellar orbits gradually align their apocenters, leading to the emergence of a coherent bar structure \citep{2023MNRAS.523.5823B}. The timescale of this formation also influences the final properties of the bar, where slowly forming bars tend to be more stable and long-lived \citep{2025ApJ...979..166W}.


As bars evolve, their properties can be further shaped by additional dynamical interactions. For instance, the bar pattern speed is influenced by the exchange of angular momentum between the disc, the bulge, and the dark matter halo \citep{1991MNRAS.250..161L, 2003MNRAS.341.1179A} or by the passage of a companion \citep{1990A&A...230...37G, 1991A&A...245L...5S, 1993A&A...280..105S}. Moreover, the presence of gas in the host galaxy can modulate this exchange, potentially maintaining a fast-rotating bar by counteracting the expected dynamical friction \citep{2023ApJ...953..173B}.

Beyond their formation and internal dynamics, bars also play a fundamental role in shaping the evolution of their host galaxies, influencing both their structure and star formation activity over time. One of the most prominent effects is their ability to drive gas inflows toward the galactic centre. This inflow can be particularly enhanced during the bar formation phase \citep{2015MNRAS.454.3641F}, or in systems with nested bars, where a secondary (inner) bar channels gas even more efficiently \citep{2023ApJ...958...77L}.

While bars play a key role in shaping the internal evolution of galaxies, their own formation and long-term survival are also influenced by external factors such as the galactic environment and cosmological context. Cosmological simulations have shown that the fraction of barred galaxies evolves over time, generally decreasing with increasing redshift \citep{2024A&A...684A.179R, 2025MNRAS.538.1587F}. Although galactic interactions, such as mergers or tidal encounters, can lead to the weakening or destruction of bars, other studies indicate that their formation is primarily governed by the intrinsic properties of the host galaxy, including disc stability and mass distribution \citep{2019MNRAS.483.2721P, 2022MNRAS.514.1006I, 2024MNRAS.529..979L, 2024A&A...684A.179R, 2025A&A...697A.236L}. Furthermore, even in a fully cosmological and environmentally rich framework, the presence of gas remains a critical factor: high gas fractions tend to suppress the formation of strong, long-lived bars by increasing turbulence and weakening disc instabilities \citep{2025ApJ...978...37A}.

Several methods have been used to identify and characterize bars in $N$-body simulations. The most common approaches rely on Fourier decomposition of the density distribution, particularly through the $m = 2$ mode amplitude $A_2$ \citep[e.g.][]{2024ApJ...965...77C}. Other methods identify bar particles through orbital alignment \citep{2016MNRAS.463.1952P} or frequency analysis aimed at detecting resonant orbital families supporting the bar \citep[e.g.][]{2016ApJ...818..141V,2023MNRAS.525.3162V}. However, most of these approaches do not directly isolate the population of particles sustaining the bar or identifies fast changes in its population throughout its evolution. Therefore, a fully dynamical characterization linking bar-supporting orbits with the three-dimensional structural evolution of the bar remains limited.

In this work, we introduce a new dynamical method to identify bar-supporting particles in $N$-body simulations. Our approach is based on spectral dynamics, originally developed by \citet{1982ApJ...252..308B, 1984MNRAS.206..159B}, and relies on the frequency analysis of individual particle trajectories to classify orbital families.

For each particle, we calculate the Fourier transform of its coordinate time series and extract the dominant frequencies and their respective amplitude that characterize its motion. Spectral methods have proven highly effective in both two- and three-dimensional potentials \citep[e.g.,][]{1998MNRAS.298....1C, 2025RMxAA..61...99S}, and have more recently been applied to $N$-body simulations to trace the evolution of orbital families \citep{2016ApJ...818..141V, 2023MNRAS.525.3162V}. Here, we introduce a new methodology by systematically identifying the bar-supporting $x_1$ population and quantifying its contribution to the global structure of the galaxy.

Once the $x_1$ family is isolated, we derive a set of diagnostics that directly link the macroscopic properties of the bar to its underlying orbital content. These include the bar mass fraction, the time evolution of its three-dimensional semi-axes, and measures of its internal structural concentration. By following these quantities over time, we obtain a detailed dynamical characterization of bar formation, growth, and long-term stability.

This approach moves beyond purely morphological diagnostics by explicitly connecting the observed bar structure to its underlying phase-space backbone. It therefore provides a physically grounded framework for studying the assembly and evolution of barred galaxies in $N$-body simulations.

This manuscript is organized as follows. In Sec. \ref{sec:methodology}, we describe the numerical models, the frozen-potential methodology, and the orbital classification procedure used to identify the $x_1$ and $x_2$ families, together with the diagnostics adopted to characterize the bar, the bulge, and the pseudo-bulge components. In Sec. \ref{sec:results}, we present the main results of our analysis, including comparisons with alternative bar-identification methods, the evolution of angular momentum redistribution, the relation between bar growth and orbital chaoticity, and the structural evolution of the bar and central components. Finally, in Sec. \ref{sec:conclusions}, we summarize our main conclusions and discuss the implications of our results for the dynamical evolution of barred galaxies.

\section{Methodology}
\label{sec:methodology}

\subsection{Galaxy models}
\label{sec:Gal_mod} 

The N-body simulations used in this study are from \citet{2023MNRAS.525.3162V}, which were built upon \citet{2019AJ....157..175V} with higher temporal resolution ($\Delta t\approx 0.98$ Myr in \citet{2023MNRAS.525.3162V}). They introduced three models ($A\lambda 03$, $A\lambda 04$, and $A\lambda 05$) with varying disc/halo dominance. Since $A\lambda 05$ forms a bar much more slowly, we focus only on $A\lambda 03$ and $A\lambda 04$ in our analysis. These isolated models initially consist of an NFW dark matter halo \citep{1996ApJ...462..563N,1997ApJ...490..493N} and an axisymmetric disc that develops a bar due to internal dynamical instabilities \citep{2002ApJ...569L..83A,2003MNRAS.341.1179A}. The models feature an exponential stellar disc, $\Sigma(R) = \Sigma_0 e^{-R/R_d}$, where $\Sigma_0 = M_d / (2\pi R_d)$. The disc vertical mass distribution follows an isothermal sheet with a constant vertical scale length $z_0$, resulting in a three-dimensional disc stellar density of \hbox{$\rho_d (R,z) = \Sigma(R) \text{sech}^2 [z / (2 z_0)] / (2 z_0)  $}.

The models are based on an equilibrium N-body realization \citep{1999MNRAS.307..162S} with $7 \times 10^6$ particles, consisting of $2 \times 10^6$ for the disc and $5 \times 10^6$ for the halo. The halo has a mass of $5.11 \times 10^{11} M_\odot$ and a concentration of 8.0, while the disc has a mass of $2.55 \times 10^{10} M_\odot$ (corresponding to a disc-to-halo mass ratio of $\approx 0.05$ in both models). The radial scale length varies across models: $R_d = 1.99$ kpc for $A\lambda 03$ and $R_d = 2.80$ kpc for $A\lambda 04$, with $z_0 = 0.2 R_d$ in both cases. 

\begin{figure}
	\includegraphics[width=\columnwidth]{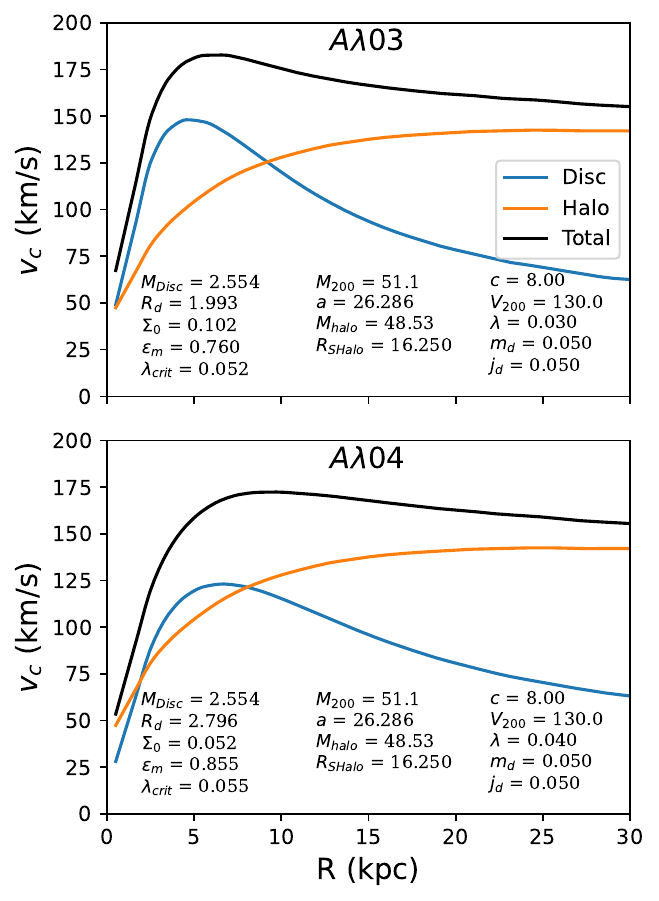}
    \caption{Initial rotation curves for $A\lambda03$ and $A\lambda04$. Their structural parameters are shown in each plot. Similar to Fig. 1 of \citet{2019AJ....157..175V}.} 
    \label{fig:InRotCur}
\end{figure}

The models were selected to represent disc-dominated systems, for which the critical spin parameter $\lambda_c$ exceeds the disc spin parameter $\lambda_d$, ensuring bar formation on relatively short timescales \citep{2023MNRAS.525.3162V}. The effects of differences in $\lambda_c$ are reflected in the components of the initial rotation curves (see Fig. \ref{fig:InRotCur}) and in the evolution of the density maps (see Fig. 1 of \citet{2023MNRAS.525.3162V}), where bar formation occurs earlier in $A\lambda03$ than in $A\lambda04$.

The collisionless N-body simulations were performed using the Gadget-2 code \citep{2001NewA....6...79S, 2005MNRAS.364.1105S}. See \citet{2019AJ....157..175V} and \citet{2023MNRAS.525.3162V} for further details.

\subsection{Bar amplitude and angle}
\label{sec:bar_amp}

We first performed a Fourier analysis of the disc particle positions to determine the bar amplitude and orientation of the snapshots. Specifically, we calculated the $m=2$ Fourier coefficients ($a_2$ and $b_2$) using particles within an annulus of width $\Delta R \approx 0.1$ kpc at radius $R$. For a given annulus, these coefficients are defined as

\begin{equation}
    \begin{split}
        a_2(R) = & \frac{1}{N(R)} \sum_{j=1}^{N(R)} \cos\left(2\phi_j\right),\\
        b_2(R) = & \frac{1}{N(R)} \sum_{j=1}^{N(R)} \sin\left(2\phi_j\right),
    \end{split}
\end{equation}

\noindent where the sum is over the $N(R)$ particles whose cylindrical radii lie within the annulus centred at $R$, and $\phi_j$ is the azimuthal angle of the $j$-th particle in the disc plane. Following \citet{2024ApJ...965...77C}, \citet{2025RMxAA..61...99S} and others, the bar amplitude and angle are given by:

\begin{equation}
    \begin{split}
        A_2(R) &=\sqrt{a_2^2+b_2^2} ; \\
       \theta_2(R) &= \arctan\left(\frac{a_2}{b_2}\right).
    \end{split}
    \label{eq:bar_amp}
\end{equation}

We applied this method to each snapshot, making $A_2$ and $\theta_2$ also time-dependent parameters: $A_2 = A_2(R,t)$ and $\theta_2 = \theta_2(R,t)$. To determine the bar angle, we calculated the average value of $\theta_2$ within the radial range of 1 to 3 kpc for each snapshot, denoted as $\bar{\theta}_2(t)$. Using $\bar{\theta}_2$, we then derived the pattern speed of the galaxy as: $\Omega_p=d\bar{\theta}_2/dt$\footnote{The derivative was calculated numerically using NumPy's {\tt gradient} function.}.

Although there are alternative methods for estimating the bar pattern speed from a single snapshot, such as the method proposed by \citet{2023MNRAS.518.2712D} and applied by \citet{2025CNSNS.14908923S}, these approaches tend to produce noisier temporal evolution compared to the method adopted in this work. Nevertheless, they can be particularly useful for simulations with limited temporal resolution.

\subsection{Frequency analysis}
\label{sec:frec_an}

In order to identify the orbital structure associated with the bar and to distinguish bar-supporting particles from those belonging to the disc, we performed a frequency analysis on both models. We used the AGAMA package \citep{2018asclsoft05008V, 2018arXiv180208255V}. Every 10 snapshots\footnote{To reduce the computational cost, the method was applied every 10 snapshots rather than at every available snapshot.} (each 9.78 Myr with our time resolution), we froze the potential by mapping the contributions from the halo and disc separately. The halo potential was determined using a multipole expansion ({\tt Multipole} function in AGAMA), while the disc potential was mapped with an azimuthal harmonic expansion ({\tt CylSpline}). These two potentials were then combined into a single potential, which was rotated in a non-inertial reference frame to follow the bar with the use of $\Omega_p$. The bar was aligned along the x-axis using $\bar{\theta}_2$.

We randomly selected 100,000 disc particles\footnote{We chose this value after performing tests using different numbers of selected disc particles from 10,000 to 2,000,000. The difference for classified $x_1$ particles, normalised to the number of selected particles, for numbers larger than 50,000, is less than 1\%. Furthermore, larger numbers of selected particles are very computational expensive for AGAMA.} from the snapshot and integrated their orbits in the frozen potential for 5 Gyrs with 1250 points (time resolution of 4 Myrs) captured at equidistant intervals\footnote{ A test in both the spatial grid resolution of the potentials and the temporal resolution of the frozen-potential simulations is presented in appendix \ref{sec:Apen_C}.}, recording their Cartesian coordinates ($x(t)$, $y(t)$, and $z(t)$). From these, we calculated the orbital radius projected onto the equatorial plane as

\begin{equation}
    R(t)=\sqrt{(x(t)-\bar{x})^2+(y(t)-\bar{y})^2}
\end{equation}

\noindent where $\bar{x}$ and $\bar{y}$ represent the average positions in the $x$ and $y$ directions, respectively. Following \citet{2025RMxAA..61...99S}, we applied a Blackman window function before performing a Fourier transform on the projected coordinates: $x(t)$, $y(t)$, and $R(t)$. To prevent artificial amplitudes at zero frequency, we first subtracted the mean value ($\bar{x}$, $\bar{y}$ and $\bar{R}$ respectively) from each coordinate. We extracted the fundamental frequencies, i.e., those with the highest amplitudes \citep{1993CeMDA..56..191L}, for these coordinates, namely $\omega_{x}$, $\omega_{y}$ and $\omega_R$. Their corresponding amplitudes, $A_{x}$, $A_{y}$, and $A_{R}$, were also recorded. 

These fundamental frequencies help characterize the shape of an orbit, as described in \citet{2025RMxAA..61...99S}. However, unlike the theoretical models in that study, where the bar is perfectly centred at the centre of mass, N-body simulations can exhibit an offset bar. If the bar is significantly displaced, the dominant frequency in $x$ or $y$ for the bar particles may be close to zero, indicating translational motion rather than intrinsic orbital behaviour. This is the reason why we subtracted the mean position from each coordinate.

Nonetheless, this approach introduces a new issue: as a result of subtracting the mean position from each coordinate, particles orbiting around the $L_4$ and $L_5$ Lagrange points may be incorrectly classified as other types of orbits. To avoid this misclassification, we filtered orbits based on their mean positions $x$ and $y$ ($\bar{x}$ and $\bar{y}$). The $L_4$ and $L_5$ Lagrange points are located near the $y$-axis, far from the bar when viewed in a rotating reference frame where the bar aligns with the $x$-axis \citep{2008gady.book.....B}. Therefore, we classify a particle as belonging to $L_4$ or $L_5$ if $\sqrt{\bar{x}^2+\bar{y}^2} > 0.5$ kpc\footnote{The threshold was chosen to exclude particles trapped around the $L_4$ and $L_5$ Lagrange points, while still retaining smily-type orbits and orbits slightly offset from the centre of mass.}. Certainly, orbits classified as $x_1$ and $x_2$ have $\sqrt{\bar{x}^2+\bar{y}^2}$ very small, near zero, because they orbit the center of the galaxy, so the chosen value of $\sqrt{\bar{x}^2+\bar{y}^2} > 0.5$ kpc ensures that the orbits are trapped around $L_4$ or $L_5$ Lagrangian points, thereby ensuring that such particles are excluded from the classification of the $x_1$ and $x_2$ orbital families.

After removing particles associated with $L_4$ and $L_5$, we encountered another source of ambiguity: chaotic orbits may display their strongest spectral peak in a way that mimics regular motion, potentially leading to misclassification. To address this issue, we introduced a quantitative measure of orbital stochasticity, which we call spectral entropy. This quantity is based on the Shannon entropy \citep{information_theory}, and evaluates how the spectral amplitude of a given coordinate is distributed across frequencies. For each orbit, the Fourier spectra of the $x$, $y$, and $z$ coordinates are calculated and their amplitudes are normalized, and the corresponding entropies quantify whether the spectrum is dominated by a small number of discrete peaks (regular motion) or spread over a broad range of frequencies (chaotic motion), following the approach of \citet{1998MNRAS.298....1C}. We denote these entropies as $S_x$, $S_y$, and $S_z$, respectively. Further details on the formulation and calculation of spectral entropy are provided in the appendix \ref{sec:Apen_A}.

\begin{figure}
	\includegraphics[width=\columnwidth]{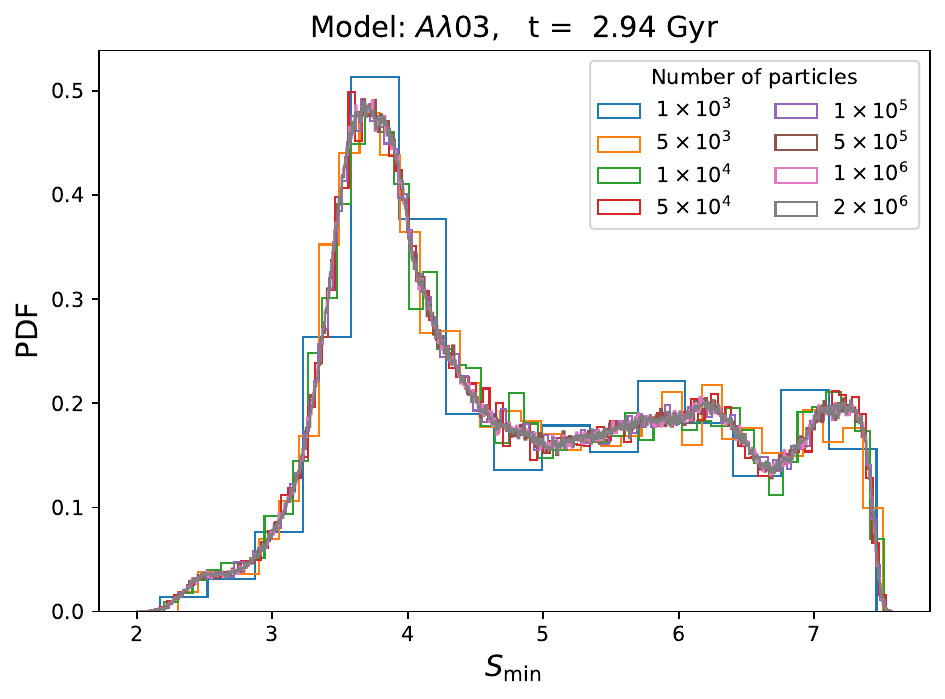}
    \caption{Distribution of $S_{\min}$ for model $A\lambda03$ at snapshot 3000 ($t=2.94$ Gyr), computed using different numbers of particles. The figure shows that the distribution has already converged for samples of $\sim10^5$ particles, with no significant variations for larger particle numbers.}
    \label{fig:conv_Smin}
\end{figure}

Using this diagnostic, we define the minimal spectral entropy ($S_{\min}$) of an orbit as the lowest value among $S_x$, $S_y$, and $S_z$. By adopting $S_{\min}$, we apply a conservative criterion in which an orbit is classified as chaotic only if stochastic behaviour is present simultaneously in all spatial coordinates, thereby reducing contamination from orbits that exhibit irregularity in a single projection due to resonances or numerical effects. Orbits with a $S_{\min}$ greater than 6.6 are classified as chaotic. This value was chosen after visually checking a large number of orbits and spectra and comparing Lyapunov indices versus $S_{\min}$. In practice, we calculated the minimal spectral entropy for the subsets of $10^5$ particles chosen before; the convergence test in Fig. \ref{fig:conv_Smin} show that increasing the number of particles does not produce significant changes in the distribution of $S_{\min}$ or in the inferred fraction of chaotic orbits.  

In a forthcoming article we will perform a quantitative study comparing the minimal spectral entropy with an independent chaos indicator, namely the Generalized Alignment Index GALI$_2$, which quantifies orbital chaos through the time evolution of the alignment of deviation vectors in phase space \citep{2001JPhA...3410029S}.  Although this threshold of $S_{\min}=6.6$  worked robustly for our simulations, it may vary depending on the total integration time, time resolution, and spectral windowing (see Appendix \ref{sec:Apen_A}).

As in \citet{2023MNRAS.525.3162V}, an orbit is considered elliptical-like if it satisfies the condition $1.9 \leq \omega_R / \omega_x \leq 2.1$, where “elliptical-like” broadly refers to orbits exhibiting elongation along either the $x$- or $y$-axis. \citet{2025RMxAA..61...99S} demonstrated that these orbits can be further subdivided into three categories: those elongated along the $x$-axis, typically associated with sticky orbits around the $x_1$ family (from now on called just $x_1$ for simplicity); those elongated along the $y$-axis, corresponding to the sticky orbits around the $x_2$ or $x_3$ families; and those that are nearly circular.

It is important to note, however, that the $x_3$ family is considerably less stable than the $x_2$ family \citep{2002MNRAS.333..861S}, and thus the fraction of sticky or chaotic orbits lingering near $x_3$ is expected to be very small. For this reason, all orbits with $y$-axis elongation falling within the elliptical-like range are classified as $x_2$ family members. \citet{2025RMxAA..61...99S} classified elliptical-like particles with $A_x / A_y > 2$ as $x_1$ orbits, and those with $A_x / A_y < 0.5$ as $x_2$ orbits. These thresholds were chosen based on simple geometric reasoning rather than statistical evidence, as the sample size at the time was insufficient for a more rigorous analysis.

\begin{figure}
	\includegraphics[width=\columnwidth]{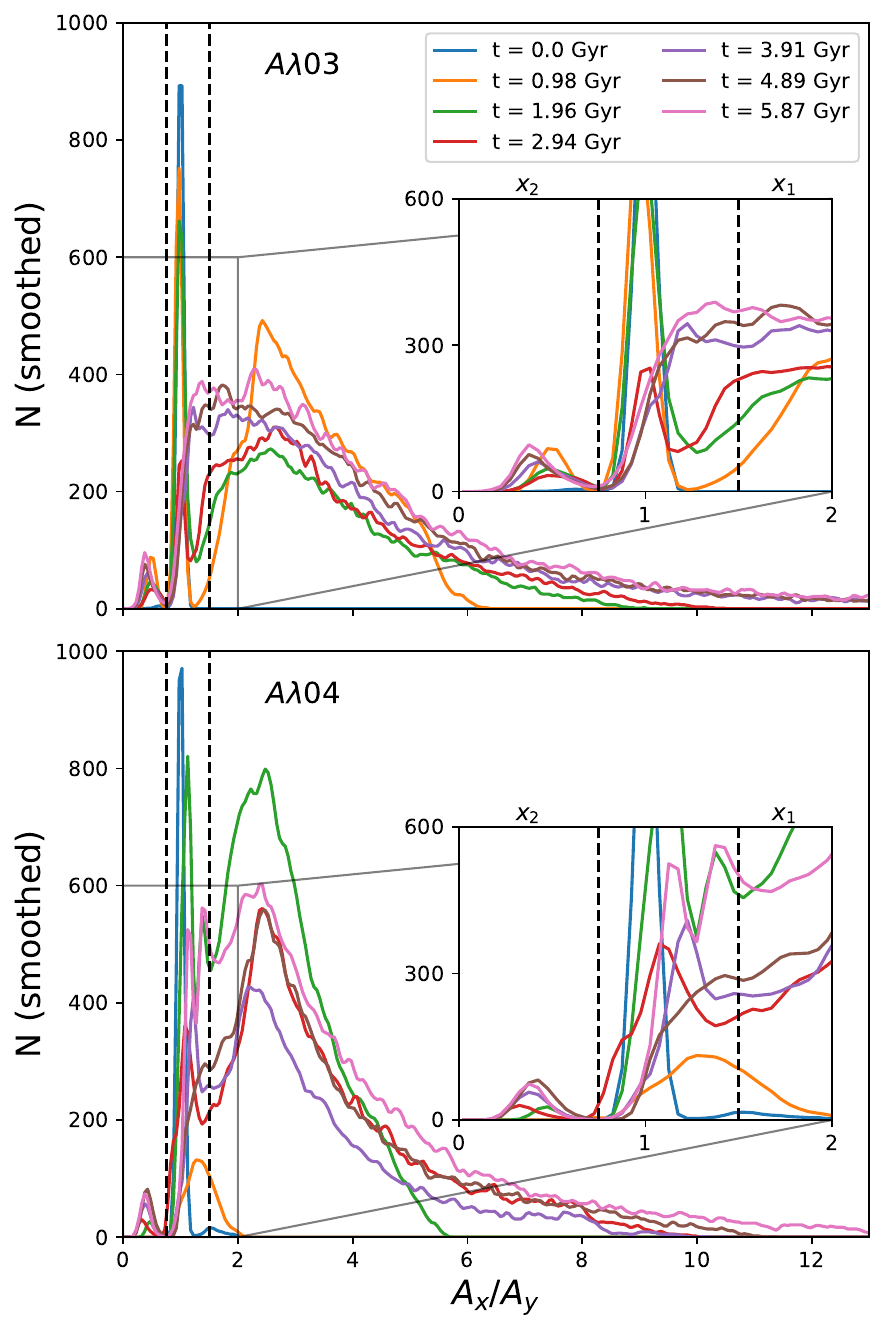}
    \caption{Distribution of amplitude ratios $A_x / A_y$ for particles satisfying the conditions $1.9 \leq \omega_R / \omega_x \leq 2.1$, $\sqrt{\bar{x}^2+\bar{y}^2}<0.5$ kpc and $S_{\min}<6.6$ in models A$\lambda$03 (top) and A$\lambda$04 (bottom). Each curve represents a different simulation time. The vertical dashed lines mark the adopted thresholds for classifying $x_2$ orbits ($A_x / A_y < 0.75$) and $x_1$ orbits ($A_x / A_y > 1.5$).}
    \label{fig:hist_amp}
\end{figure}

\begin{figure*}
	\includegraphics[width=0.8\textwidth]{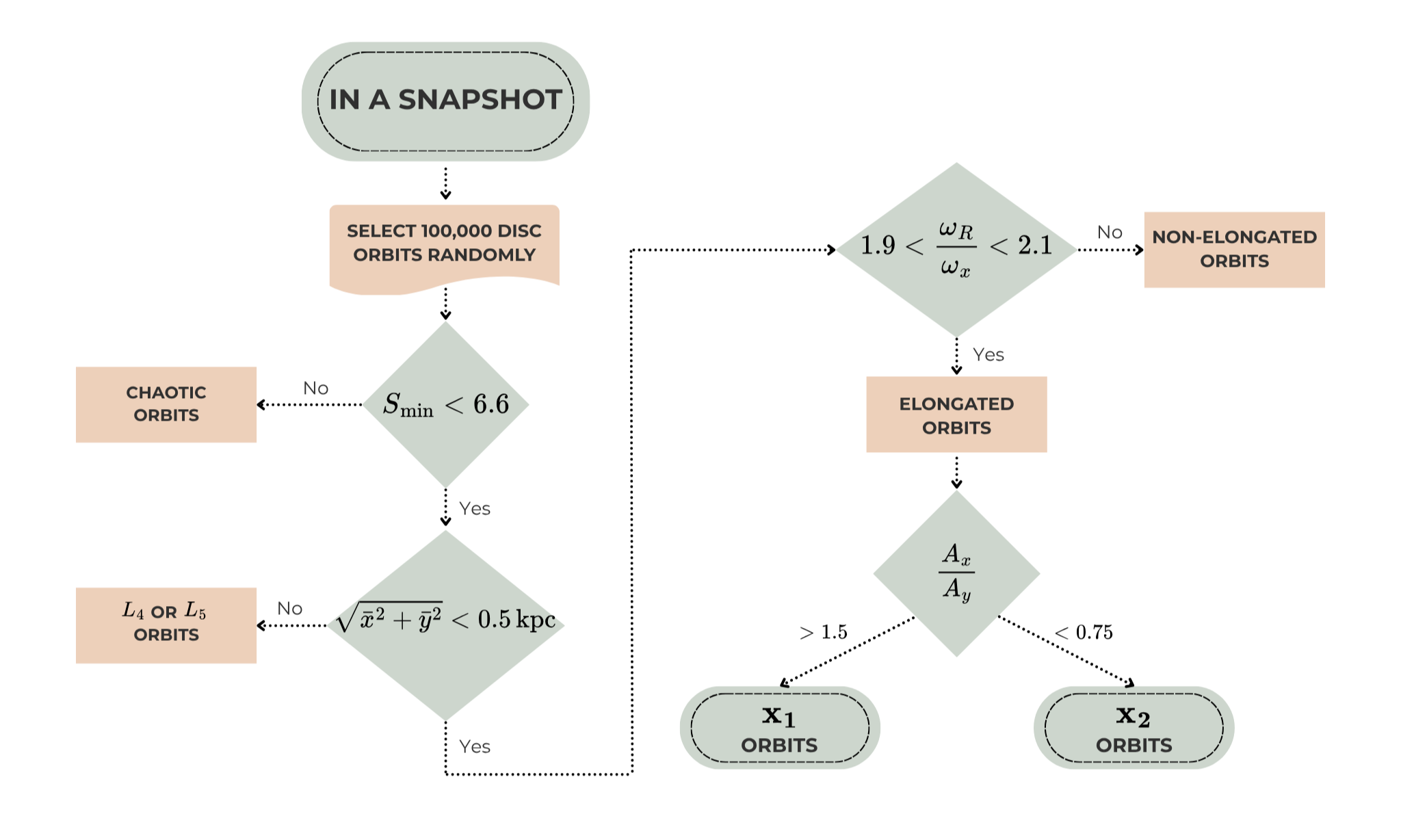}
    \caption{Flowchart summarizing the orbit-classification procedure used to identify the $x_1$ and $x_2$ families in a given snapshot.}
    \label{fig:flowchart}
\end{figure*}

In Fig. \ref{fig:hist_amp}, we show the distribution of $A_x / A_y$ for all elliptical-like particles in both models. On the $x_2$ range, the distribution exhibits a well-defined separation, with a clear minimum near $A_x / A_y \approx 0.75$ (indicated by the left black vertical dashed line in Fig. \ref{fig:hist_amp}). In contrast, the $x_1$ side does not display a sharp or consistent minimum. To address this, we examined the two-dimensional distributions of $x_1$ orbits in both models (see Appendix \ref{sec:Apen_B}), which led us to adopt here $A_x / A_y = 1.5$ (right black vertical dashed line in Fig. \ref{fig:hist_amp}) as a reasonable threshold for the $x_1$ family. This choice ensures a consistent and practical classification of elliptical-like orbits into the $x_1$ and $x_2$ families while excluding elliptical orbits that are located outside, rather than within, the bar region (see Fig. \ref{fig:2D_ra_L40}). For greater clarity, Figure \ref{fig:flowchart} presents a flowchart summarizing the procedure followed in this work to identify orbits belonging to the $x_1$ and $x_2$ families.

\begin{figure}
	\includegraphics[width=\columnwidth]{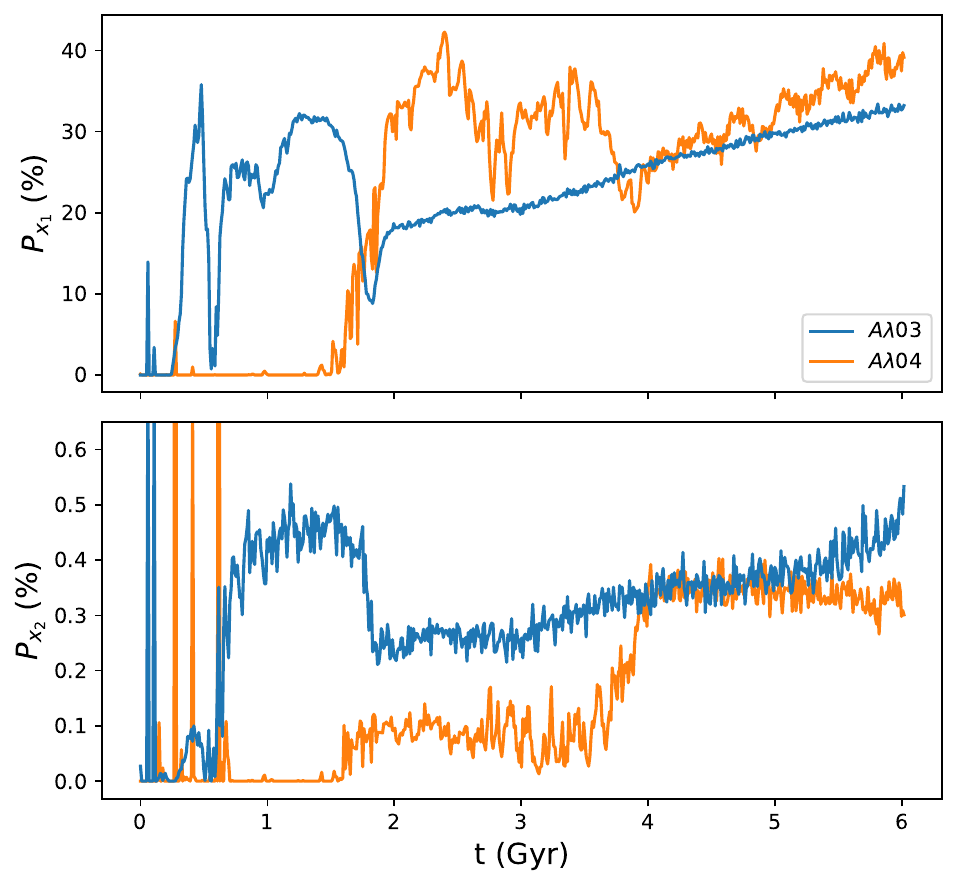}
    \caption{Evolution of the proportion of $x_1$ (top panel) and $x_2$ (bottom panel) family members for both models.}
    \label{fig:px1_px2}
\end{figure}

With the orbits belonging to the $x_1$ and $x_2$ families identified, we quantify their proportion out of the 100,000 orbits as $P_{x_1}$ and $P_{x_2}$. As described in Sec. \ref{sec:bar_amp}, this process is repeated for each snapshot, making these quantities time-dependent. The top panel of Fig. \ref{fig:px1_px2} shows the evolution of $P_{x_1}$ for both models, while the bottom panel displays the evolution of $P_{x_2}$. Although their specific trends will be discussed in detail in Secs. \ref{sec:results}, this figure provides the general context for how the relative mass fraction of the $x_1$ and $x_2$ families evolves throughout the simulation, where the $x_1$ orbits correspond to the main bar-supporting family, while the $x_2$ orbits are linked to secondary or inner bars seen in both simulations \citep[among others]{contopoulos1980orbits, athanassoula1992, skokos2002, Martinez_Valpuesta_2006} and observations  \citep{1993A&A...277...27F, 2002AJ....124...65E, 2011MSAIS..18..145E, 2015ApJS..217...32B}.

\subsection{Diagnostics for bar structure}
\label{sec:Diagnostics}

\begin{figure*}
  \includegraphics[width=\textwidth]{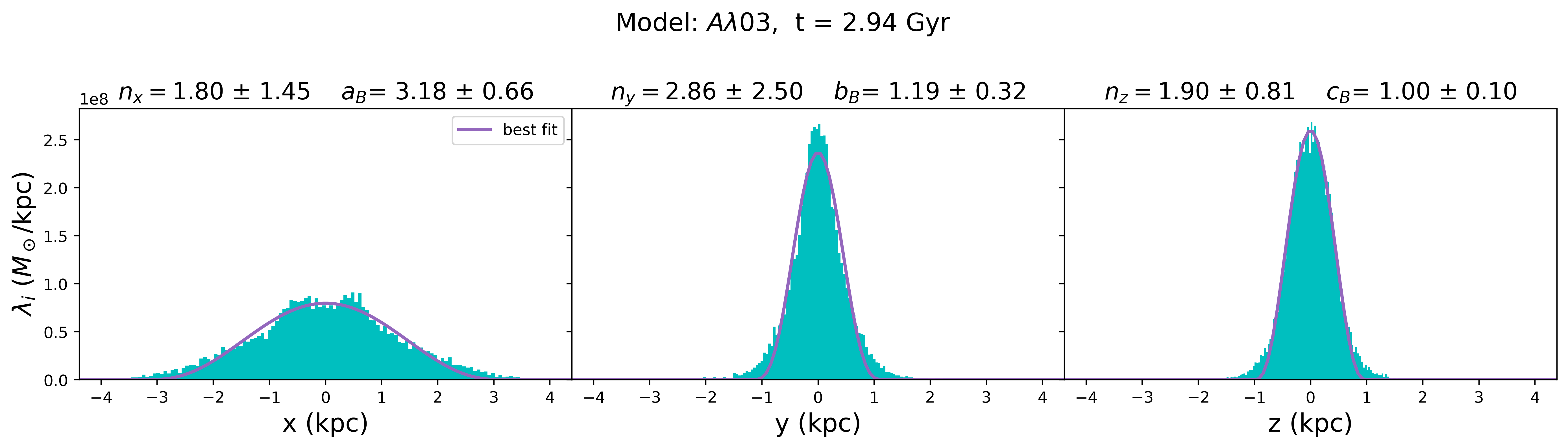}
  \caption{Histograms of the linear density distribution ($\lambda_i$) of $x_1$ orbits along the $x$-axis (left), $y$-axis (middle), and $z$-axis (right) for a snapshot of model $A\lambda03$. The solid curves show the best-fit Ferrers bar profile for each axis. The corresponding semi-axis lengths and steepness indices are indicated above each panel.}
  \label{fig:lambda}
\end{figure*}

Once the bar members are identified, we can proceed to determine its properties. Considering the bar as an ellipsoid similar to Ferrers bars \citep{1877QJPAM..14....1F}, we can statistically estimate its semi-axes and the steepness along each axis. To do this, we assume the bar behaves similarly to a Ferrers bar and calculate its linear density, $\lambda$, along each axis. For a Ferrers bar, the linear density along the $x_i$ axis (where the indices $i$, $j$ and $k$ can take values from 1 to 3 with $i \ne j \ne k$ ) is given by:

\begin{align}
\lambda_i &= \int_{-\infty}^\infty \int_{-\infty}^\infty 
\rho(x_i, x_j, x_k)\, dx_j\, dx_k \nonumber \\[1.2ex]
&= \begin{cases}
\displaystyle \frac{105}{32} \frac{M_B}{a_i(n+1)}
\left( 1 - \frac{x_i^2}{a_i^2} \right)^{n+1}, & |x_i| < a_i, \\[1.2ex]
0, & \text{otherwise.}
\end{cases}
\label{eq:dens_lin}
\end{align}

\noindent where $a_i$ represents the semi-axis along $x_i$, $M_B$ is the total bar mass, and $n$ is an index that characterizes the steepness of the density profile. In standard Ferrers bar models, the same value of $n$ is assumed for all three axes. However, from Eq. \ref{eq:dens_lin}, we note that $\lambda_i$ depends only on quantities associated with the $x_i$ axis. This allows us to assign a distinct steepness index to each axis, which we denote as $n_x$, $n_y$, and $n_z$. Fig. \ref{fig:lambda} shows the linear density distribution of $x_1$ orbits for a snapshot of model $A\lambda03$ at $t = 2.94$ Gyr, along with the best-fit Ferrers profile for each axis. The corresponding semi-axes ($a_B$, $b_B$, and $c_B$ for $x$, $y$, and $z$, respectively), as well as the steepness indices $n$ and their fitting uncertainties, are also indicated. Since this is a statistical approach, we only evaluate the properties of the bar once a sufficiently large number of particles is available to ensure robust statistics. In practice, this condition is met when at least 5\%\footnote{The 5\% threshold is not intended as a universal definition of bar formation. It is used only as an operational lower limit to ensure that the $x_1$ population is sufficiently populated for stable estimates of the bar semi-axes and density indices.} of the orbits belong to the $x_1$ family ($P_{x_1} > 5\% =5{,}000$ particles), which allows us to construct well-sampled histograms and perform stable fits of the density profiles.

A notable feature is a small dip in the central region of the $x$-axis distribution (left panel of Fig. \ref{fig:lambda}). This dip may arise because orbits near the bar centre fail to meet the condition $A_x/A_y > 1.5$, and are therefore either chaotic or closer to circular/spherical. The presence of this dip could suggest the existence of a bulge and/or pseudo-bulge, a possibility that we will revisit later in Sec. \ref{sec:bulge_ev}.

By applying this procedure to each snapshot, we can follow the time evolution of the semi-axes ($a_B$, $b_B$, and $c_B$) and the steepness indices ($n_x$, $n_y$, and $n_z$). These quantities provide a time-dependent characterization of the bar size, shape, and internal density structure. In Sec. \ref{sec:results}, we present their temporal evolution for both models and examine how they relate to the orbital content and global bar properties.

\subsection{Bulge and pseudo-bulge identification}
\label{sec:bul-psb}

We also identified the components of the central spheroid, hereafter referred to as the bulge, and the pseudo-bulge. In this work, the term pseudo-bulge is used in a purely dynamical sense, following \citet{2022MNRAS.515.1524Z} notation, to denote a centrally concentrated stellar component identified through orbital circularity and binding energy, rather than by morphological or photometric criteria. 

In order to identify the central components, we applied the methodology outlined by \citet{2022MNRAS.515.1524Z}, with a few minor adjustments tailored to our specific case. Notably, this analysis was not conducted on all stellar particles in the simulation. Instead, it was applied to the 100,000 sampled particles in Sec. \ref{sec:frec_an}.

In addition, we calculated the gravitational potential and its derivatives using AGAMA. Since the frozen potentials had already been constructed for the orbital analysis, this approach allowed us to derive the circular velocity consistently from the radial derivative of the gravitational potential as:

\begin{equation}
    v_c=\sqrt{R\frac{d\Phi}{dR}}  
\end{equation}

\noindent where $R = \sqrt{x^2 + y^2}$, rather than using the expression $v_c = \sqrt{GM(<R)/R}$ as in \citet{2022MNRAS.515.1524Z}. This adjustment enabled us to estimate the circular angular momentum ($j_c(R,0) \equiv R v_c(R,0)$) and the total energy ($E_c \equiv \frac{1}{2} v_c^2 + \Phi$) in the galactic plane. To mitigate the effects of a non-axisymmetric particle distribution, these quantities were averaged over four positions in the plane: $(R, 0, 0)$, $(0, R, 0)$, $(-R, 0, 0)$, and $(0, -R, 0)$. The procedure was applied at 100 logarithmically spaced radii spanning the distance from the innermost to the outermost stellar particle in the galaxy.

As in the reference study, we used these profiles to parametrize $j_c(E_c)$, which was then used to interpolate the circular angular momentum over the energy distribution of the sample particles. Except for the modifications described above, we followed the methodology of \citet{2022MNRAS.515.1524Z} to determine the particle circularity, $\eta \equiv j_z/j_c$, and the energy cut-off, $E_{\rm cut}$, which separates the more bound from the less bound particles.

Once $E_{\rm cut}$ had been determined, we excluded the particles belonging to the $x_1$ and $x_2$ orbital families from the subsequent decomposition. This step is necessary because bar particles can otherwise be misclassified as part of the central components, as noted by \citet{2022MNRAS.515.1524Z}. After removing the bar particles, we followed the MORDOR decomposition to identify the bulge and pseudo-bulge components.

\begin{figure}
	\includegraphics[width=\columnwidth]{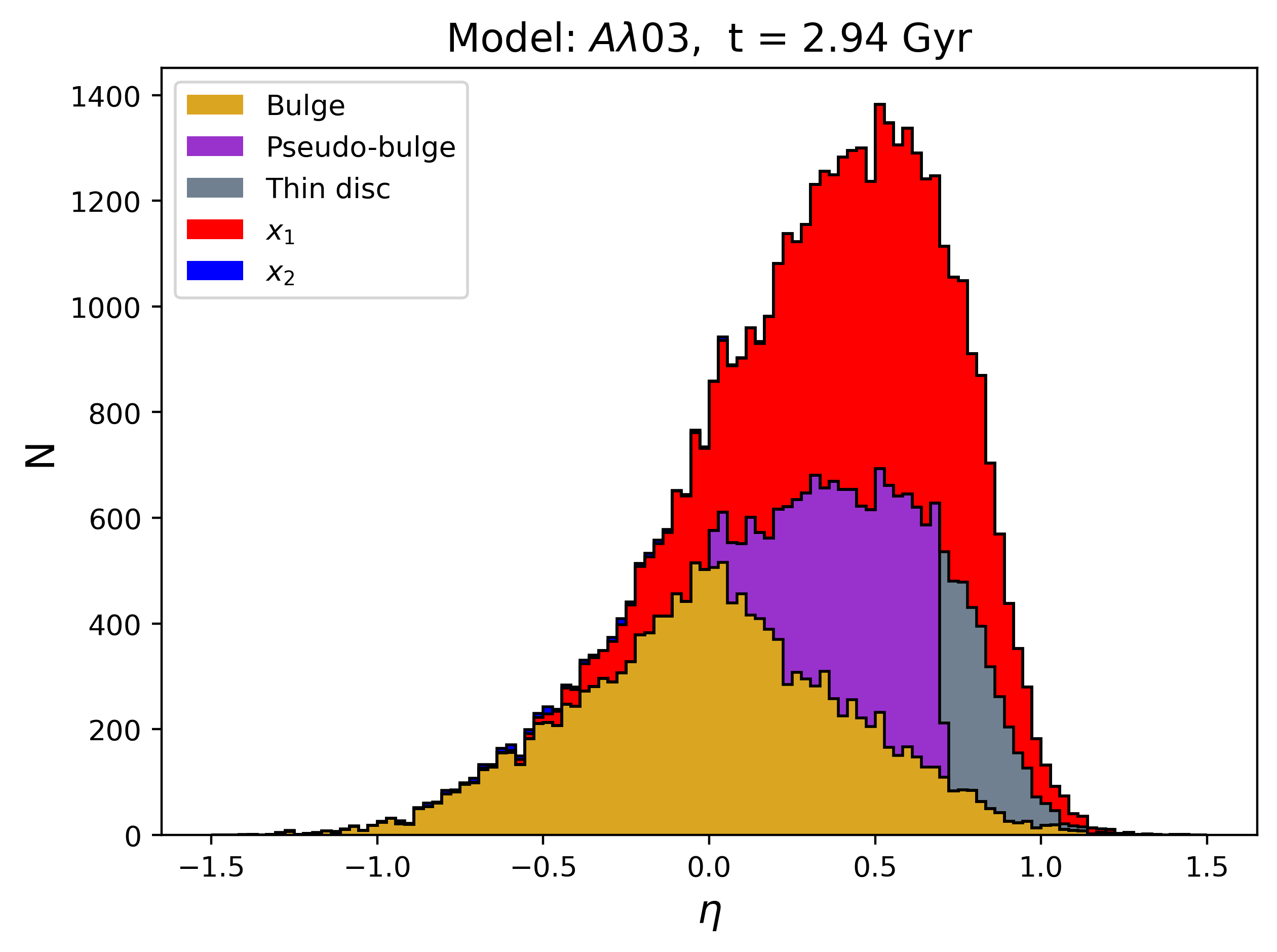}
    \caption{Circularity ($\eta$) distribution of all selected particles with energies $E \leq E_{\rm cut}$ for a snapshot of model $A\lambda03$ at $t = 2.94$ Gyr. The histogram is decomposed into the thin disc (gray), pseudo-bulge (purple), bulge (gold), $x_1$ family (red), and $x_2$ family (blue), with the components stacked for visual clarity.}
    \label{fig:circularity}
\end{figure}

Figure \ref{fig:circularity} presents the $\eta$ distribution for all the sample particles selected in Sec. \ref{sec:frec_an} with $E < E_{\rm cut}$ of one snapshot of the $A\lambda03$ model, distinguishing between the five components with such energy: bulge, pseudo-bulge, thin disc and the $x_1$ and $x_2$ orbital families.

\section{Results}
\label{sec:results}

Having established the numerical setup and the diagnostics used to characterize orbital structure, we now present the results of our analysis. In this section, we examine the bar morphology, its structural properties, the temporal evolution of chaoticity, and the interplay between the bar and the other galactic components, for both models $A\lambda03$ and $A\lambda04$. Our aim is twofold: first, to assess the consistency of our results with previous studies; and second, to demonstrate the additional insight provided by our methodology in linking orbital dynamics with the global structural evolution of the bar.

\subsection{Comparison with alternative approaches}
\label{sec:px1_vs_A2}

\subsubsection{Validation against Fourier bar's amplitude}

As extensively discussed in previous studies \citep{contopoulos1980orbits, athanassoula1992, skokos2002}, the $x_1$ family of periodic orbits constitutes the dynamical backbone of rotating galactic bars. These orbits are elongated along the bar major axis and remain stable within the bar-supporting region, thereby providing the primary orbital framework of the barred structure. Particles belonging to the $x_1$ family can therefore be regarded as genuine bar members.

\begin{figure}
  \includegraphics[width=\columnwidth]{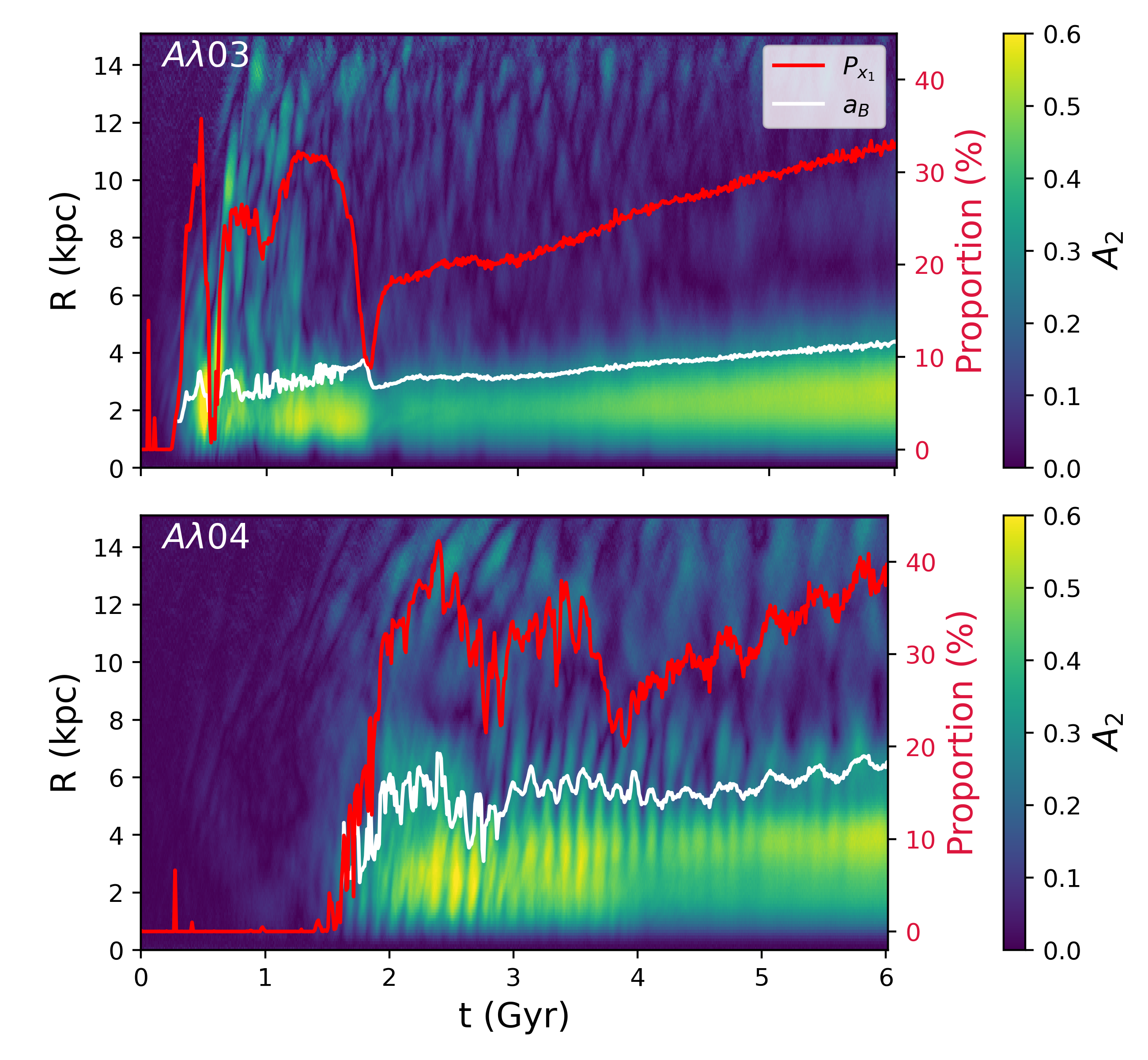}%
  \caption{Comparison between the bar amplitude, $A_2(R,t)$ (represented by the color map), the proportion of orbits belonging to the $x_1$ family, $P_{x_1}(t)$ (depicted by the red line, corresponding to the right vertical axis), and the major semi-axis of the bar $a_B(t)$ (white line, corresponding to the left vertical axis), for models A$\lambda$03 (top panel) and A$\lambda$04 (bottom panel).}
  \label{fig:A2_vs_Px1}
\end{figure}

To validate our orbital classification, we compare the fraction of $x_1$ orbits, $P_{x_1}$, and the bar major semi-axis, $a_B$, with the radial distribution of the Fourier bar amplitude $A_2(R,t)$ (Eq. \ref{eq:bar_amp}). Fig. \ref{fig:A2_vs_Px1} shows this comparison for both models. A clear correspondence is observed: the temporal evolution of $P_{x_1}$ closely follows the strength of the $A_2$ signal within the bar region, while the semi-major axis $a_B$ matches the radial extent over which $A_2$ is maximal. 

Nevertheless, $A_2$ is often reported as a single one-dimensional measure obtained by averaging over a given radial range. Following this common practice, we calculated the mean bar amplitude as

$$
\bar{A}_2(t) = \langle A_2(1 < R < 3\,\mathrm{kpc},\, t) \rangle,
$$

\noindent in analogy with the definition of $\bar{\theta}_2$ in Sec. \ref{sec:bar_amp}. The evolution of both $\bar{A}_2$ and $P_{x_1}$ from the frozen potential method is shown in Fig. \ref{fig:comp_method}.

Although derived from fundamentally different approaches (orbital classification in our case and Fourier decomposition of the surface density for $A_2$) both diagnostics yield consistent descriptions of the bar structure and its evolution. This agreement supports the interpretation that the identified $x_1$ particles correspond to the dynamically trapped orbits that physically constitute the bar.

Nevertheless, the two quantities differ conceptually. The frozen-potential method has a direct dynamical interpretation: it measures the fraction of particles (and therefore the mass fraction) trapped in the bar-supporting orbital family. In contrast, $\bar{A}_2$ is a Fourier-based diagnostic that quantifies the strength of the $m=2$ mode without explicitly identifying the underlying orbital structure. 

A further distinction concerns dimensionality. From $A_2(R,t)$ one extracts essentially one-dimensional radial information, such as an estimate of the bar length. By contrast, the frozen-potential classification allows a fully three-dimensional characterization of the bar (see Sec. \ref{sec:Diagnostics}), including the evolution of the semi-axes and shape parameters. The orbital approach therefore provides additional structural and dynamical insight beyond the information encoded in Fourier amplitudes alone.

\subsubsection{Time-resolved orbital methods}

In addition to the frozen-potential analysis, we considered two alternative methods that use the original particle trajectories.

First, following \citet{2023MNRAS.525.3162V}, we performed a frequency analysis of orbit segments extracted directly from the simulation in the reference frame co-rotating with the bar. For each selected time, orbital parameters were calculated over a 2 Gyr window centered on that time, and orbits were classified according to the criteria described in Sec. \ref{sec:frec_an}. The entropy threshold distinguishing regular from chaotic orbits was set to $S_{\min}=5.5$, reflecting the dependence of spectral entropy on both time resolution (0.98\,Myr) and integration length (2\,Gyr). Because the original trajectories are already available, this approach is computationally less expensive and was applied to all $2{,}000{,}000$ disc particles.

Second, we implemented the apsidal-alignment method proposed by \citet{2016MNRAS.463.1952P}. In this framework, bar membership is determined from the alignment of orbital apsides with the bar major axis. For each of all $2{,}000{,}000$ disc orbits, we calculated the mean absolute angular offset over 20 azimuthal periods, $\langle \delta\theta_{||} \rangle_{20}$, which measures how closely the outer turning point remains aligned with the bar. Small values indicate bar-supporting behaviour, while large values correspond to misaligned or non-bar orbits.

Bar particles were identified using two approaches: (i) a $k$-means clustering algorithm with $k=2$, selecting the cluster whose centroid satisfied $\langle \delta\theta_{||} \rangle_{20} < \pi/8$, and (ii) a direct threshold criterion, classifying particles as bar members whenever $\langle \delta\theta_{||} \rangle_{20} < \pi/8$.

\begin{figure}
  \includegraphics[width=\columnwidth]{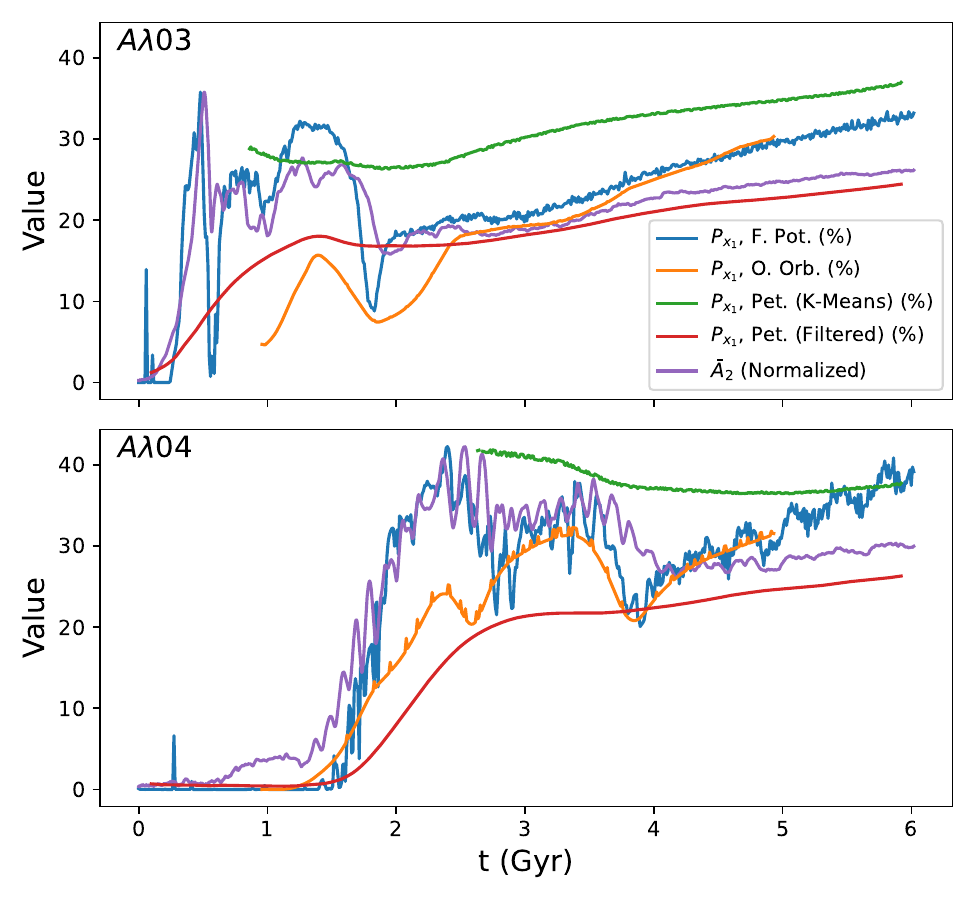}
  \caption{Comparison of five bar diagnostics for models $A\lambda03$ (top panel) and $A\lambda04$ (bottom panel). 
  Shown as a function of time are: the fraction of $x_1$ orbits ($P_{x_1}$) calculated in the frozen potential (blue), the fraction of $x_1$ orbits obtained from the original orbits over 2 Gyr intervals (orange), the fraction of bar particles identified using the apsidal-alignment method of \citet{2016MNRAS.463.1952P} via $k$-means clustering (green) and via a fixed threshold $\langle \delta\theta_{||} \rangle_{20} < \pi/8$ (red), and the normalized bar amplitude $\bar{A}_2$ (purple). 
  All quantities are expressed as percentages except for $\bar{A}_2$, which is normalized for visual comparison.}
  \label{fig:comp_method}
\end{figure}

Figure~\ref{fig:comp_method} compares the five diagnostics for both models. The frozen-potential $P_{x_1}$ closely tracks the normalized $\bar{A}_2$, reinforcing the consistency between our method and one of the most used methods to quantify the bar amplitude in the literature.

By contrast, the bar fractions derived from both the original-orbit and apsidal-alignment methods exhibit noticeably smoother temporal evolution. This behaviour reflects the intrinsic time-averaging built into these techniques: the original-orbit method averages over 2\,Gyr windows, while the apsidal-alignment approach evaluates alignment over 20 azimuthal periods. In both cases, successive measurements rely on overlapping orbital segments, naturally suppressing short-term fluctuations. Additionally, the original-orbit method does not span the full temporal range, since each measurement requires data both before and after the selected time.

\subsubsection{Clustering versus threshold identification}

Among all diagnostics, the apsidal-alignment method implemented via $k$-means clustering shows the largest deviation from the others. At early times, the inferred bar fraction appears truncated, as the clustering algorithm does not always identify a centroid satisfying $\langle \delta\theta_{||} \rangle_{20} < \pi/8$. In these stages, the separation between aligned and non-aligned populations is not sufficiently pronounced for a clear two-cluster partition.

Moreover, when a bar cluster is identified, the corresponding bar fractions are systematically higher than those obtained with the other methods. This discrepancy likely reflects the relatively weak contrast between the underlying $\langle \delta\theta_{||} \rangle_{20}$ populations in these models. When aligned and non-aligned orbits overlap significantly in parameter space, the unsupervised clustering may assign a broader set of particles to the “bar” group, thereby inflating the inferred fraction.

In such regimes, the fixed-threshold criterion provides a more stable and physically transparent classification than the clustering approach.

\subsubsection{Strengths and limitations of the frozen-potential method}

One might argue that the original-orbit or apsidal-alignment methods offer a more direct physical interpretation, since they rely exclusively on actual particle trajectories. However, the frozen-potential approach also has a clear dynamical meaning: it determines whether an orbit is trapped within the bar’s potential well, which is effectively equivalent to being dynamically part of the bar. Furthermore, by adjusting the integration time per snapshot, the frozen-potential method can achieve higher frequency resolution than the original-orbit analysis.

The primary limitation of the frozen-potential technique is its computational cost. In addition to evolving the full $N$-body simulation, it requires integrating frozen-potential orbits at each analysed snapshot. This imposes a trade-off between particle sampling and temporal resolution; in practice, we analysed $100{,}000$ particles and one snapshot out of every ten.

A second limitation concerns the determination of the pattern speed $\Omega_p$, which is required to transform to the rotating frame. Any inaccuracy in $\Omega_p$ directly propagates into the orbital classification. The spurious early-time peaks in $P_{x_1}$ and $P_{x_2}$ (Fig.~\ref{fig:px1_px2}) arise because our estimate of $\Omega_p$ is based on $\bar{\theta}_2$, which becomes reliable only once a coherent bar is present. Consequently, measurements prior to full bar formation are more uncertain.

Despite these challenges, the frozen-potential method provides a uniquely powerful dynamical framework for studying barred galaxies. Unlike Fourier- or alignment-based diagnostics, it enables a direct decomposition of phase space into orbital families and quantifies their relative contributions to the bar. This approach not only identifies the bar-supporting $x_1$ population, but also allows a detailed characterization of its three-dimensional structure, stability, and temporal evolution. 

By linking the global properties of the bar to its underlying orbital content, the frozen-potential analysis offers a physically grounded perspective on bar assembly that cannot be obtained from purely morphological measures. For this reason, it forms the foundation of the dynamical interpretation presented in the following sections.

\subsection{Angular momentum}
\label{sec:ang_mom}

While the role of bars in redistributing angular momentum is well established \citep{Athanassoula2003, sellwood2014}, the specific pathways through which angular momentum is exchanged within the disc remain an active area of research \citep{10.1093mnrasstz2824, 2023ApJ...942..106J, Trapp2024}. To investigate this, we analysed the angular momentum content of the galactic disc in our models in order to clarify how the bar modifies the orbital structure. For each snapshot, we calculated the angular momentum along the $z$–axis for every particle, $L_{z,i}$, and summed over all particles within annuli of width $\Delta R \approx 75$ pc at radius $R$. Repeating this procedure across snapshots yields the global distribution of angular momentum as a function of radius and time, $L_z(R,t)$, following the same annulus-based approach described in Sec.~\ref{sec:bar_amp} for $A_2(R,t)$ and $\theta_2(R,t)$.

In addition, following a similar method by \citet{2023ApJ...942..106J}, we calculated the change in angular momentum relative to the initial disc value as:

\begin{equation}
    \Delta L_z(R, t)=\frac{L_z(R,t)-L_z(R,0)}{L_z(R,0)}.
\end{equation}

\begin{figure*}
  \includegraphics[width=\textwidth]{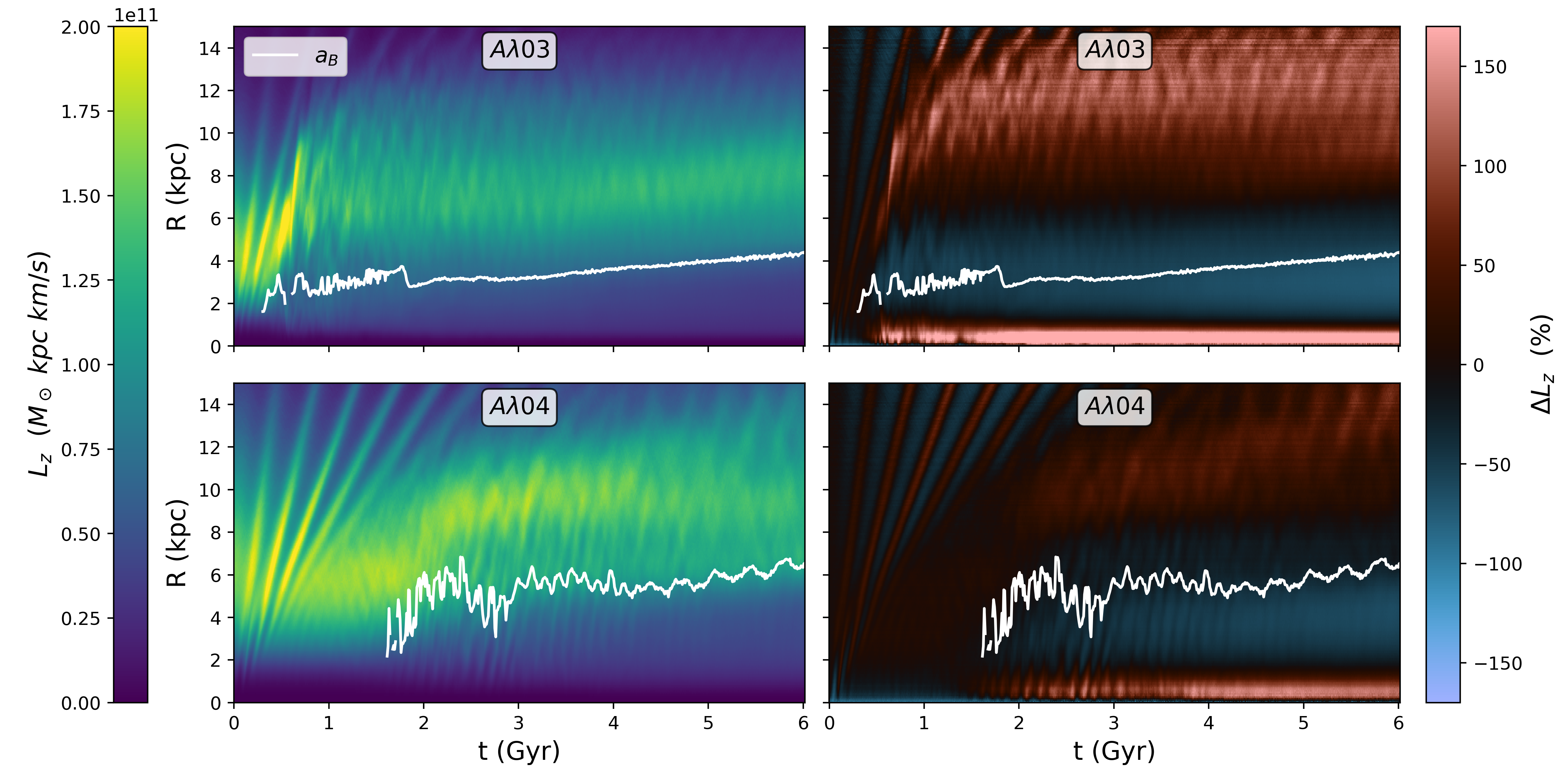}
  \caption{Temporal and radial distribution of angular momentum for models $A\lambda03$ (top) and $A\lambda04$ (bottom). The left panels show the absolute angular momentum $L_z(R,t)$, while the right panels display the relative change with respect to the initial distribution, $\Delta L_z(R,t)$. In all panels, the white curve marks the bar semi-major axis $a_B$.}
  \label{fig:Lz_vs_aB}
\end{figure*}

Figure~\ref{fig:Lz_vs_aB} presents colormaps of both $L_z(R,t)$ and $\Delta L_z(R,t)$ for models $A\lambda03$ and $A\lambda04$, with the bar semi-major axis $a_B$ overplotted to facilitate direct comparison between the bar and the angular momentum distribution. In the left panels, where the evolution of $L_z(R,t)$ is shown, it is noticeable that at the beginning of the simulations, “waves” of angular momentum transfer propagate outward from the inner galaxy, gradually weakening over time. And, once the bar is established, it dominates the redistribution of angular momentum toward the outer disc.

Another important feature is the apparent role of $a_B$ as a boundary between regions of relatively high and low angular momentum. The bar region is consistently characterized by lower angular momentum. At first glance, this might be attributed to the smaller radii of the bar particles. However, the right panels of Fig.~\ref{fig:Lz_vs_aB}, which show $\Delta L_z(R,t)$, demonstrate that the bar region actually loses angular momentum relative to its initial state. This loss occurs only after bar formation, indicating that the bar itself is responsible for the angular momentum depletion in its vicinity.

From the right panels of Fig.~\ref{fig:Lz_vs_aB}, we also note a region just outside the bar where particles continue to lose angular momentum, although less efficiently than those within the bar itself. This suggests that this outer region may act as a reservoir of material feeding the bar. As these particles lose angular momentum, their orbital velocities decrease and their guiding radii shrink; being located near the bar, they are subsequently incorporated into its structure.

\begin{figure}
  \includegraphics[width=\columnwidth]{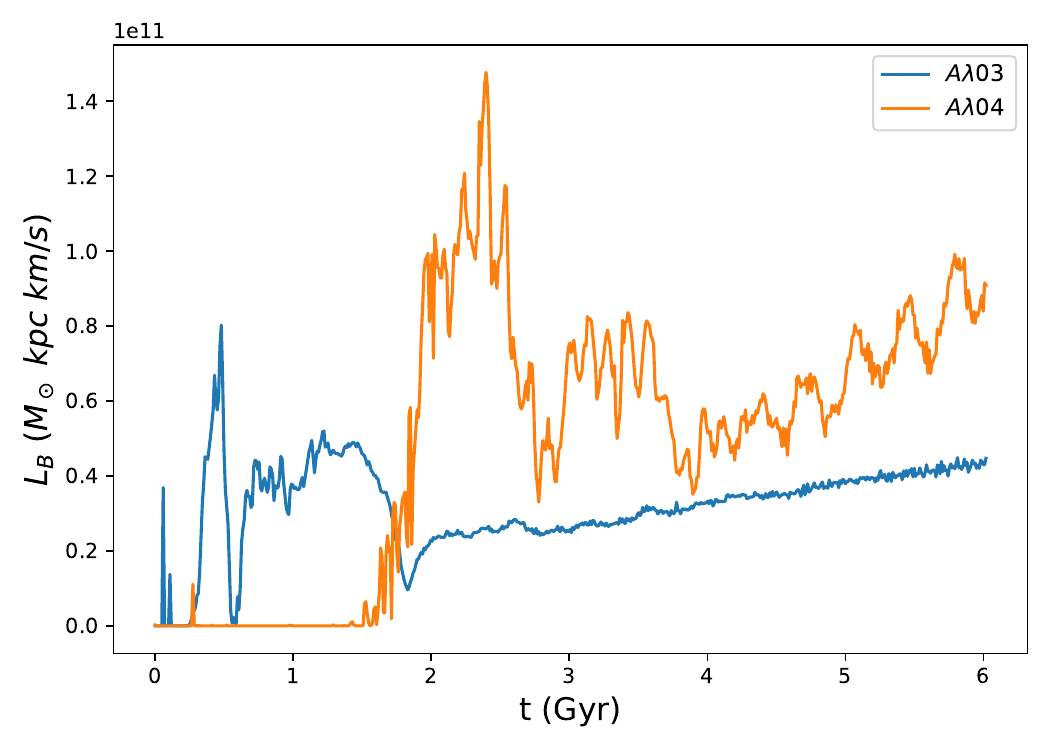}%
  \caption{Evolution of the angular momentum on the z-axis ($L_B$) for both models.}
  \label{fig:L_B}
\end{figure}

If we calculate the angular momentum of the bar (or at least a fraction of it, since it is applied to the 100,000 particles sample) as:

\begin{equation}
    L_B=\sum_{i\in B}L_{z,i},
\end{equation}

\noindent where $B$ denotes the set of bar particles at a given time, we can trace its temporal evolution (Fig.~\ref{fig:L_B}). The evolution of $L_B$ closely follows that of $P_{x_1}$ in both models, indicating that the bar’s angular momentum primarily increases through the accretion of particles. In contrast, changes in the orbital speeds or radii of existing bar particles appear to contribute much less significantly to the overall growth of $L_B$. This reinforces the picture in which the bar grows by trapping nearby particles rather than by significantly altering the kinematics of its initial population. In this way, the bar acts both as a sink of angular momentum and as a driver of its outward transfer, a dual role that underpins its long-term growth and impact on galactic structure. 

This dual role of the bar as both a sink and a driver of angular momentum transfer is consistent with the findings of \citet{10.1093mnrasstz2824}, who performed a more detailed analysis of the underlying mechanisms of angular momentum exchange through direct torque measurements and orbital decomposition. Despite the differing levels of analysis, both studies highlight the bar as the dominant structure governing the long-term redistribution of angular momentum within galaxies.

In summary, our analysis supports the classical picture in which bars drive the secular redistribution of angular momentum within galaxies. The results presented here show that the bar can act as a self-regulated dynamical structure whose role as both a sink and a source of angular momentum depends on the relative contribution of chaotic and regular orbital families. 

This behaviour is in qualitative agreement with previous theoretical and numerical studies (\citealt{Athanassoula2003, 2003MNRAS.341.1179A, Athanassoula2013, Trapp2024}). At the same time, the consistency between our results and these established findings provides an independent validation of the methodology employed in this work. In particular, the ability to link angular momentum exchange to the bar identified through our analysis demonstrates that the method reliably captures the dynamical processes governing the evolution of barred galaxies.

Having established that our approach yields results consistent with other validated methods, we now explore the additional insights made possible by the direct identification of bar particles. This allows us to investigate in greater detail the evolution of the bar and its relation to other galactic components.

\subsection{Evolution of the bar and chaos}
\label{sec:chaos}

If we examine the top panel of Fig. \ref{fig:px1_px2} or Fig. \ref{fig:A2_vs_Px1}, we can identify three distinct evolutionary phases of $P_{x_1}$ in both models. The first phase shows a rapid rise immediately after bar formation, followed by an abrupt decline, reflecting a substantial loss of $x_1$ particles whose dynamical origin will be discussed below. This is followed by an intermediate stage of reactivation, during which bar orbits attempt to reorganize, but again end with a marked reduction in the $x_1$ population. Finally, in the third phase, $P_{x_1}$ enters a more gradual and sustained growth that persists until the end of the simulation. For a better understanding, the three phases are shown for both models in Fig. \ref{fig:prop_comp} in the next section.

\begin{figure}
  \includegraphics[width=\columnwidth]{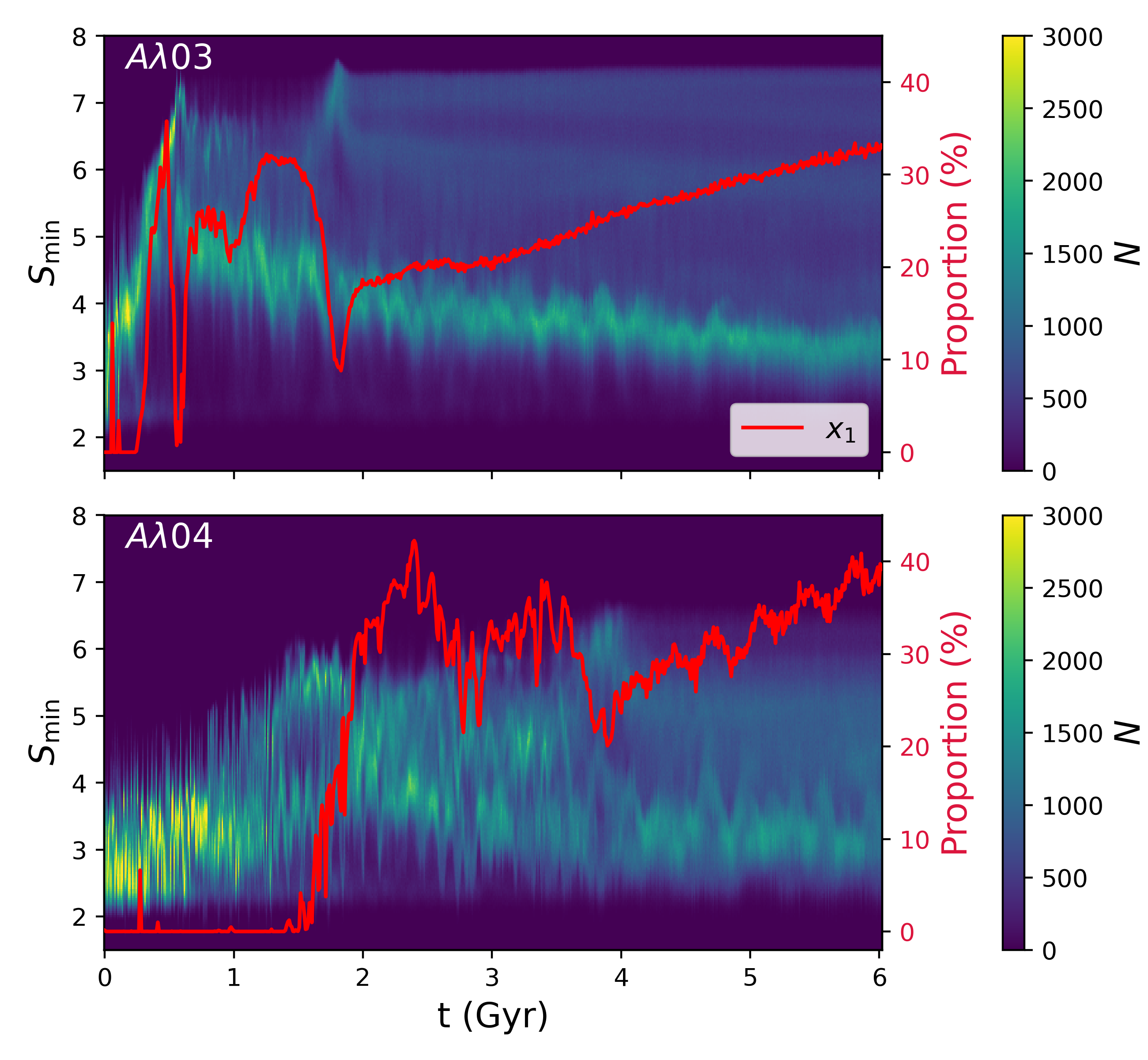}%
  \caption{2D histograms showing the distribution of minimal spectral entropy ($S_{\min}$) over time, alongside the temporal evolution of the relative proportion of the $x_1$ family orbit, shown as a red line and referenced to the right vertical axis, for models A$\lambda$03 (top panel) and A$\lambda$04 (bottom panel). As in Fig. \ref{fig:prop_comp}, a Hampel filter was applied to the time evolution of all components to remove outliers.}
  \label{fig:Smax_vs_Px1}
\end{figure}

This behaviour may be related to the chaotic nature of the orbits. Using the minimal spectral entropy ($S_{\min}$, see Section \ref{sec:frec_an} and Appendix \ref{sec:Apen_A}), we can distinguish between regular orbits (characterized by low $S_{\min}$ values) and chaotic ones (high $S_{\min}$). In this framework, larger values of $S_{\min}$ indicate more chaotic orbital behaviour. Fig. \ref{fig:Smax_vs_Px1} shows the evolution of $S_{\min}$ distributions for the 100,000 calculated orbits in each snapshot. For comparison, the evolution of the proportion of $x_1$ family orbits is also included in the same figure, allowing a direct comparison between the bar component and the degree of orbital chaos.

In Fig. \ref{fig:Smax_vs_Px1}, in both cases, a strong correlation is evident between the fraction of $x_1$ orbits and the overall degree of chaos, as indicated by the $S_{\min}$ of the orbits. During the initial phase of bar growth, there is a marked increase in the fraction of orbits with high $S_{\min}$ values, indicating enhanced chaotic behaviour. This coincides with a steep rise in $P_{x_1}$, which is then interrupted by a sharp decline. The drop in $P_{x_1}$ aligns with transient bursts of orbital chaoticity, as reflected by the elevated $S_{\min}$ distribution (more pronounced in $A\lambda03$ than in $A\lambda04$).

After the initial drop in $P_{x_1}$, both models enter into a reorganization episode. During this interval, $P_{x_1}$ rises again, concurrently, the $S_{\min}$ distribution shifts toward lower values and becomes narrower, indicating a net reduction in chaoticity, although a high-$S_{\min}$ tail persists for a subset of particles. The end of this phase is marked by a second decline in $P_{x_1}$, which coincides with renewed bursts of high $S_{\min}$ in part of the population, suggesting that some bar particles became chaotic and escaped from the bar potential. 

Following this second decline, both models enter a third, secular stage in which $P_{x_1}$ grows steadily until the end of the simulation. This phase indicates a progressive and stable reorganization of the orbital structure in favour of the $x_1$ family. Concurrently, most of the $S_{\min}$ distribution shifts toward lower values, reflecting a reduction in chaotic motion and a more stable dynamical configuration. Nevertheless, in both models a persistent high-$S_{\min}$ tail remains from the end of the second evolutionary stage through to the end of the simulation.

Overall, these results indicate that the evolution of the bar is closely linked to the temporal redistribution of orbital chaos. Periods of enhanced chaoticity coincide with abrupt reductions in the fraction of $x_1$ orbits, suggesting that chaotic diffusion weakens orbital trapping within the bar region. Conversely, phases characterized by a narrowing of the $S_{\min}$ distribution and a reduction in high-entropy orbits are associated with sustained growth of the $x_1$ population, indicating a stabilization of the bar-supporting orbital structure.

In this sense, the bar does not grow monotonically, but rather through successive episodes of dynamical reorganization, during which chaotic orbits are either trapped into, or released from, the $x_1$ family. The long-term secular increase in $P_{x_1}$ is therefore accompanied by a global reduction in chaotic behaviour, reflecting the emergence of a progressively more stable barred configuration.

\subsection{Bulge and pseudo-bulge identification}
\label{sec:bulge_ev}

\begin{figure*}
  \includegraphics[width=\textwidth]{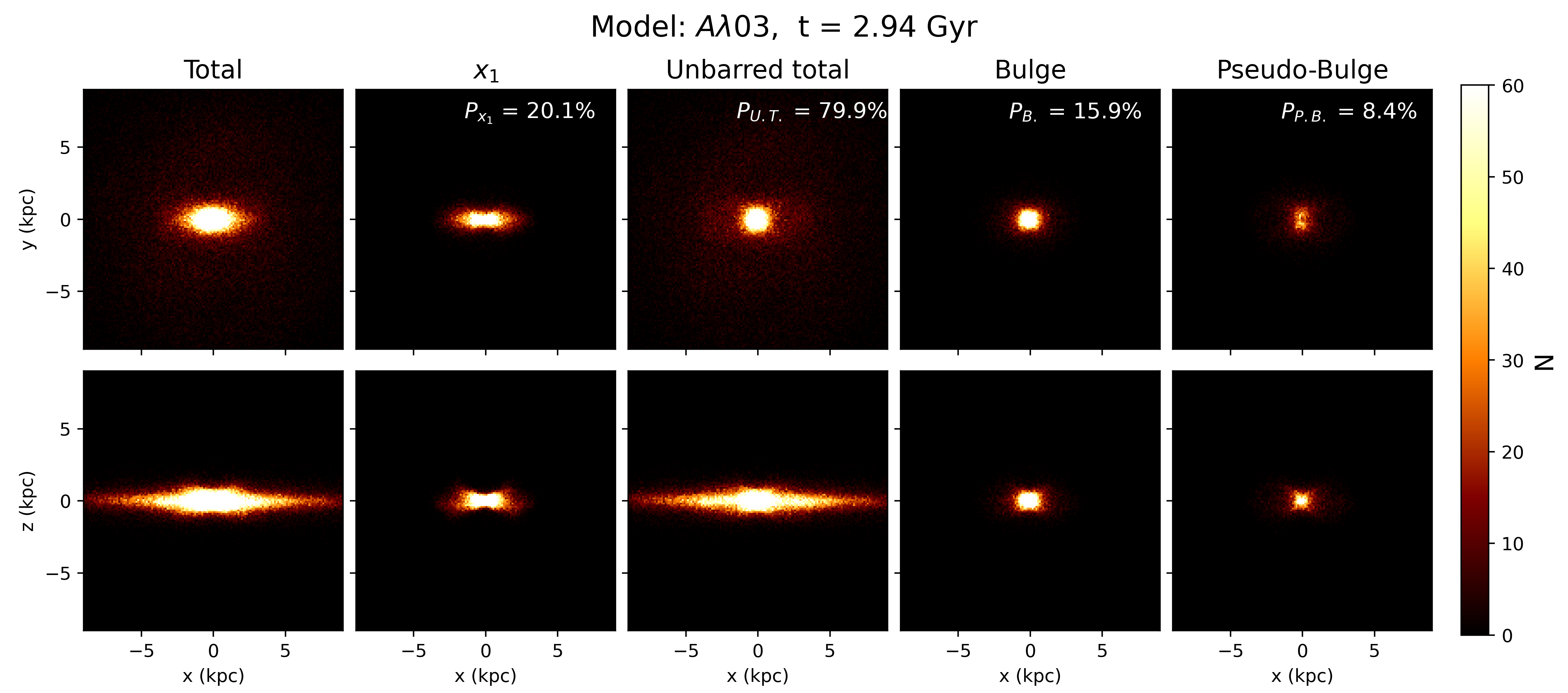}
  \caption{Face-on (top row) and edge-on (bottom row) 2D histograms showing the spatial distribution of a 100,000 particle sample from model $A\lambda03$ at $t = 2.94$ Gyr. The decomposition illustrates the contributions of distinct stellar components: the total stellar distribution (first column), the bar (second), the total stellar particles excluding only the bar (third), the classical bulge (fourth), and the pseudo-bulge (fifth). The colour scale indicates the projected particle density. The proportion of particles with respect to the total disc particles are shown in each of their corresponding top panels.}
  \label{fig:componentes_L30}
\end{figure*}

As mentioned earlier, the left panel of Fig. \ref{fig:lambda} shows a small dip in the central region, which is also apparent in the second column of Fig. \ref{fig:componentes_L30}. At first glance, this might appear to be an isolated feature. However, this dip is present in a large number of snapshots from both models. Interestingly, a similar feature was also observed by \citet{2023ApJ...953..173B}, who used a kinematic method to decompose the disc. 

\begin{figure*}
  \includegraphics[width=\textwidth]{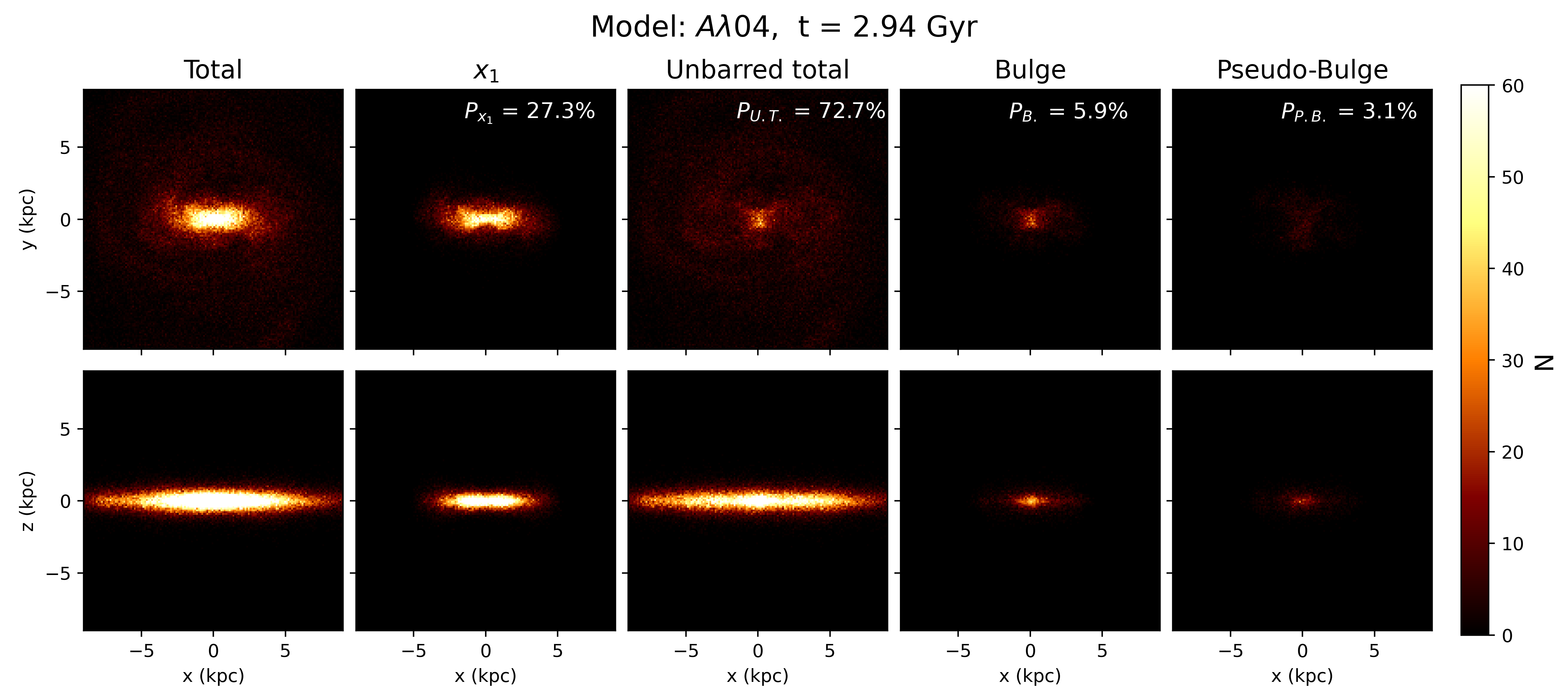}
  \caption{Same as Fig. \ref{fig:componentes_L30} but for $A\lambda 04$.}
  \label{fig:componentes_L40}
\end{figure*}

As explained in Sec. \ref{sec:Diagnostics}, the orbits near the centre of the bar struggle to meet the condition $A_x/A_y>1.5$ to be counted as part of the bar, meaning the shapes of their orbits are either chaotic or more circular/spherical. Hence, this dip could be explained by a significant number of particles belonging to the bulge or pseudo-bulge components rather than the bar. This interpretation is further supported by the third column of Figures \ref{fig:componentes_L30} and \ref{fig:componentes_L40}, which show the spatial distribution of stellar disc particles not associated with the bar. In there, we clearly see a prominent central structure that likely corresponds to the bulge or pseudo-bulge.

To determine whether the prominent central structure corresponds to the bulge or the pseudo-bulge, we identified the particles associated with these components following the procedure described in Sec. \ref{sec:bul-psb}, applied to both models and to each analysed snapshot. Figures \ref{fig:componentes_L30} and \ref{fig:componentes_L40} show 2D distributions of the three components studied, displayed in both face-on and edge-on projections for both $A\lambda03$ and $A\lambda04$. From Figures \ref{fig:componentes_L30} and \ref{fig:componentes_L40}, it is qualitatively evident that the central dip in the bar component (second column) is largely filled by particles belonging to the bulge or pseudo-bulge (fourth and fifth columns respectively). This qualitative result suggests that the central dip in the bar is not an isolated artifact, but rather a natural outcome of the contribution from bulge and pseudo-bulge particles.

\begin{figure}
  \includegraphics[width=\columnwidth]{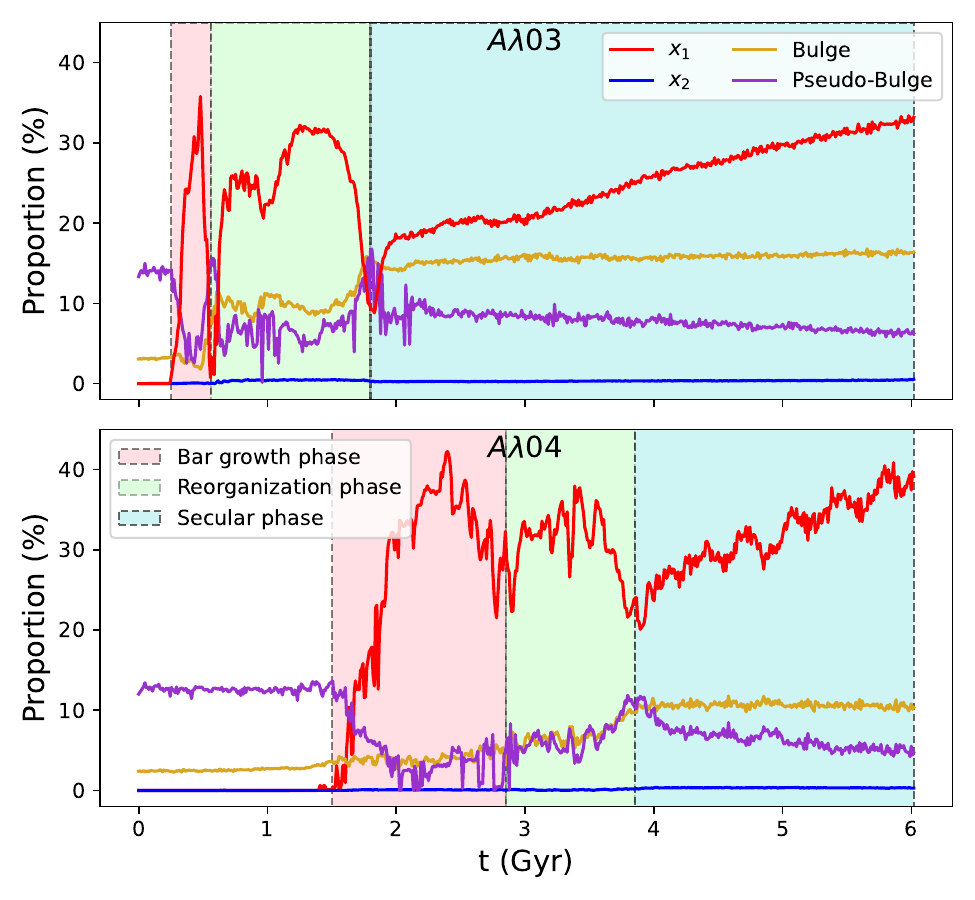}%
  \caption{Evolution of the relative proportion of four galactic components ($x_1$, $x_2$, bulge, and pseudo-bulge) with respect to the total number of particles in the sample. To reduce noise, particularly in the bulge and pseudo-bulge, a Hampel filter was applied to the time evolution of all components to remove outliers. The three evolutionary phases of the bar are also shown with different background colours.}
  \label{fig:prop_comp}
\end{figure}

In Fig. \ref{fig:prop_comp}, we examine the temporal evolution of the relative contributions of all components to quantify how their proportions change throughout the simulation. A notable feature is present during all the simulation for both models: the fraction of bar particles ($x_1$ orbits) is almost anti-correlated with that of the pseudo-bulge. When the bar fraction increases, the pseudo-bulge fraction decreases, and vice versa. This behaviour suggests that a significant portion of bar particles originates from the pseudo-bulge, and particles that cease to belong to the bar tend to be reassigned to the pseudo-bulge. This interpretation is consistent with the idea of the pseudo-bulge being strongly interconnected to the bar \citep{2011MNRAS.415.3308G, 2013ApJ...772...36G}, making it a reservoir from which the bar can both gain and lose particles.

Another interesting feature of Fig. \ref{fig:prop_comp} is the presence of bulge and pseudo-bulge particles at the beginning of the simulations, even though the initial conditions contain only a stellar disc and a dark matter halo (Sec. \ref{sec:Gal_mod}). This does not imply that these components are initially present. Instead, because the MORDOR decomposition is purely kinematic, the particles are classified according to their orbital properties. The initial bulge and pseudo-bulge fractions therefore reflect the kinematic state of the stellar distribution prior to bar formation, rather than the structural components adopted to initialize the simulations.

\subsubsection{Impact of bar-particle identification on bulge and pseudo-bulge decomposition}

As discussed by \citet{2022MNRAS.515.1524Z}, the MORDOR algorithm does not explicitly identify bar particles. Consequently, orbits supporting the bar may be misclassified as belonging either to the bulge or to the pseudo-bulge component. In our analysis, we address this limitation by explicitly removing particles associated with the $x_1$ and $x_2$ orbital families after determining $E_{\rm cut}$ but before identifying those particles belonging to the bulge and pseudo-bulge.

\begin{figure}
  \includegraphics[width=\columnwidth]{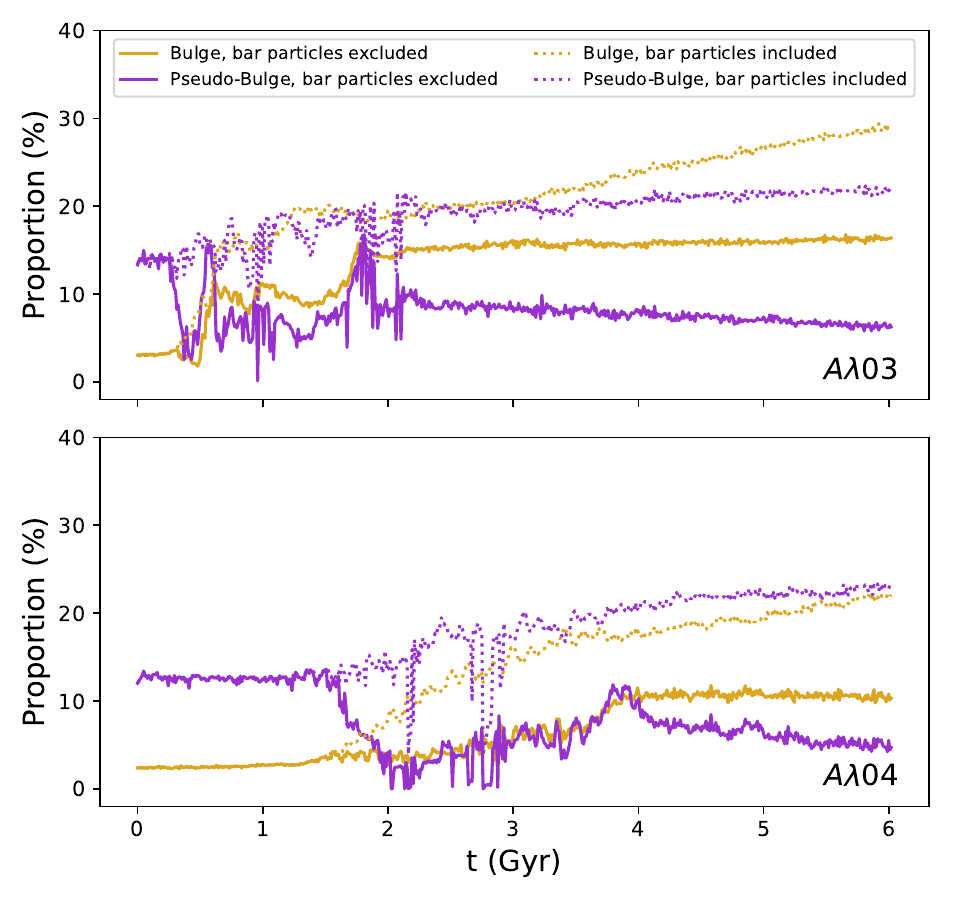}%
  \caption{Temporal evolution of the bulge and pseudo-bulge mass fractions in models $A\lambda03$ (top) and $A\lambda04$ (bottom). Solid lines correspond to the MORDOR decomposition after removing particles belonging to the $x_1$ and $x_2$ bar-supporting families, while dotted lines show the results when bar particles are included in the classification. To reduce noise, a Hampel filter was applied to the time evolution of all components to remove outliers.}
  \label{fig:comp_mordor}
\end{figure}

To quantify the impact of this correction, Fig.~\ref{fig:comp_mordor} compares the temporal evolution of the bulge and pseudo-bulge mass fractions obtained when bar particles are included in the MORDOR classification (dotted lines) and when they are excluded (solid lines).
In both models, the inclusion of bar particles leads to a significant overestimation of the bulge-related components. The bulge mass fraction can increase by nearly a factor of two at late times when bar particles are not removed. The pseudo-bulge is even more strongly affected, reaching nearly a factor of three at late times in both simulations when bar particles are retained.

The differences shown in Fig.~\ref{fig:comp_mordor} highlight the importance of explicitly identifying bar-supported orbits prior to structural decomposition. Without this correction, part of the bar mass is artificially reassigned to spheroidal components, potentially biasing interpretations of bulge growth and secular evolution. By isolating bar particles through their orbital properties, our method enables a cleaner separation between dynamically distinct components, leading to a more physically consistent characterization of the inner galaxy structure.

\subsection{Analysis of the bar properties}
\label{sec:an_prop}

The evolution of the bar cannot be fully characterized by the fraction of $x_1$ family members alone. As described in Sec. \ref{sec:Diagnostics}, we also quantified the bar geometry through the determination of its semi-axes ($a_B$, $b_B$, and $c_B$) and the steepness of its density profile along each axis via the corresponding indices ($n_x$, $n_y$, and $n_z$) by fitting the semi-axis and $n$ index in Eq. \ref{eq:dens_lin} in each axis.

\begin{figure}
  \includegraphics[width=\columnwidth]{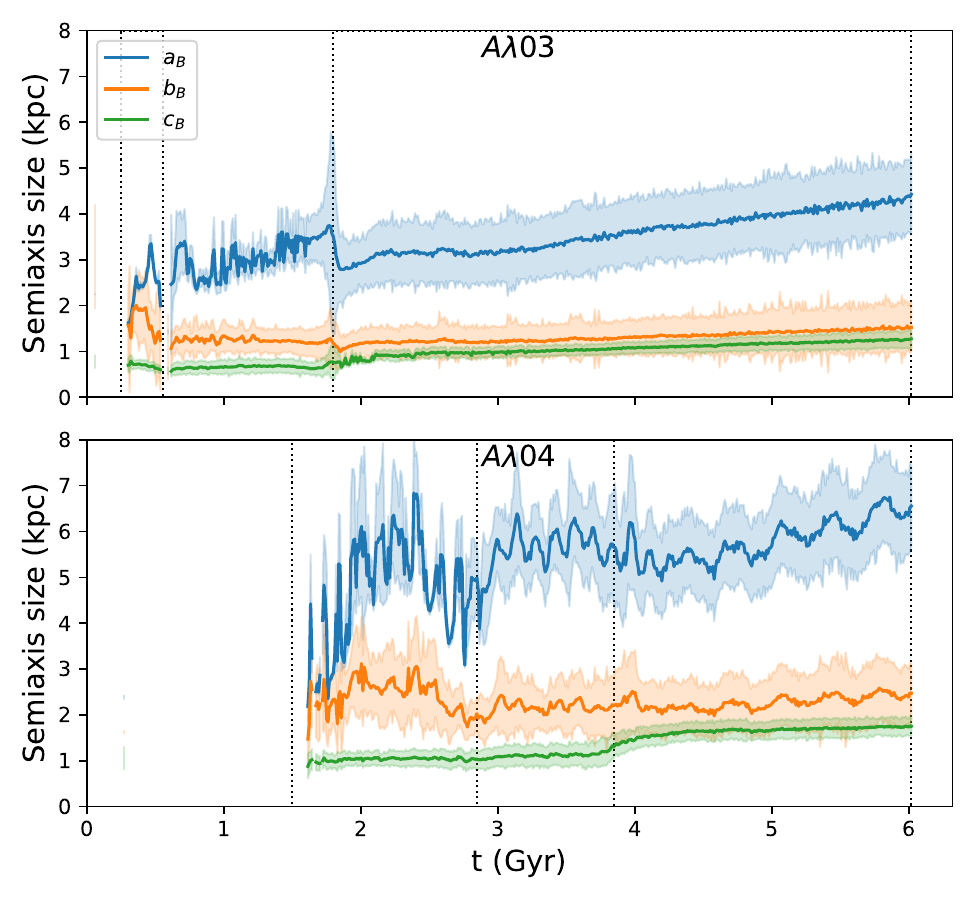}%
  \caption{Evolution of the bar semi-axes ($a_B$, $b_B$, and $c_B$) for models A$\lambda$03 (top) and A$\lambda$04 (bottom). The semi-axis lengths are measured only after the bar contains more than 5\% of the disc particles. Shaded regions indicate the fitting uncertainties derived as described in Sec. \ref{sec:Diagnostics}. Vertical dotted lines mark the boundaries between the three evolutionary phases.}
  \label{fig:semiaxes}
\end{figure}

\begin{figure}
  \includegraphics[width=\columnwidth]{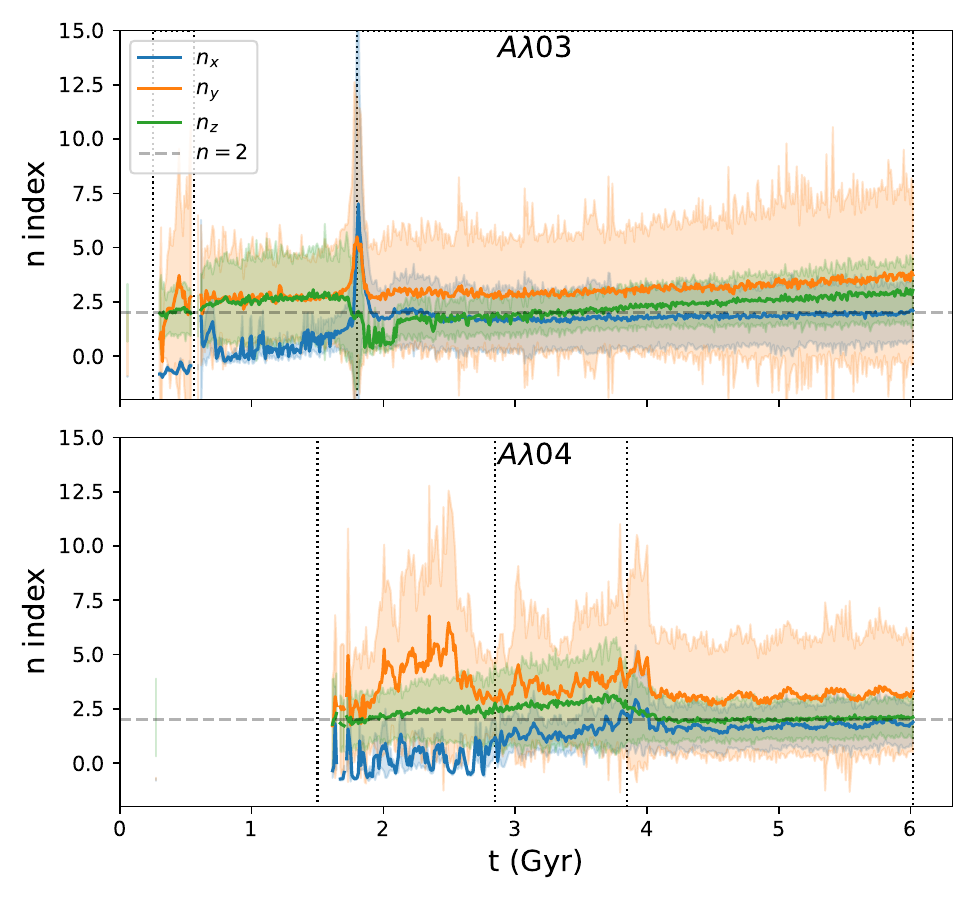}
  \caption{Time evolution of the bar $n$ indices ($n_x$, $n_y$, and $n_z$) in models A$\lambda$03 (top panel) and A$\lambda$04 (bottom panel). The semi-axis steepnesses are measured only after the bar contains more than 5\% of the disc particles. The shaded bands represent the uncertainties associated with the fitted indices calculated in Sec. \ref{sec:Diagnostics}. The horizontal dashed line marks the canonical Ferrers value, $n=2$, commonly adopted in analytical bar potential models. Vertical dotted lines indicate the transitions between the three evolutionary phases identified in each model.}
  \label{fig:index_ev}
\end{figure}

Applying this procedure to each snapshot allows us to follow the temporal evolution of both the bar size and its internal structural profile. Figures~\ref{fig:semiaxes} and \ref{fig:index_ev} show the evolution of the semi-axes and the $n$ indices, respectively, including the associated fitting uncertainties, for both models calculated in Sec. \ref{sec:Diagnostics}.

\subsubsection{Evolution of the bar's semi-axes}

From Fig.~\ref{fig:semiaxes} two important aspects of the bar evolution become evident. First, our method does not only provide a reliable estimate of the bar length, represented by the major semi-axis $a_B$ (as already suggested by the comparison with $A_2$ in Fig.~\ref{fig:A2_vs_Px1}), but also allows the determination of the intermediate and minor semi-axes, $b_B$ and $c_B$. This provides a more complete geometric description of the bar, enabling a multidimensional characterization of its structure.

Second, the three semi-axes evolve at different rates throughout the simulation. As a consequence, the axis ratios ($b_B/a_B$ and $c_B/a_B$) are not constant in time, indicating that the bar does not simply grow in size but also changes its shape as it evolves. Tracking these variations provides additional information about the structural evolution of the bar that cannot be captured when considering only its radial extent.

\subsubsection{Evolution of the bar steepnesses}

From Fig. \ref{fig:index_ev}, a simple yet more important observation can be made: $n_x \neq n_y \neq n_z$ for the majority of the simulation time. This indicates that the common assumption in analytical models of a constant and isotropic $n$ index (i.e., $n=n_x = n_y = n_z$, as in the Ferrers bar profile) is a significant simplification and not representative of the evolving structures observed in self-consistent models. Nevertheless, this assumption is understandable: analytical models prioritize tractability and computational efficiency over capturing the full complexity of dynamical evolution.

Furthermore, the frequent use of Ferrers bars with $n=2$ in the literature is likely motivated by practical considerations. \citet{1984A&A...134..373P} introduced a recursive algorithm that efficiently calculates the Ferrers potential and its derivatives for integer values of $n$. Among these, the case $n=2$ provides a favourable compromise between mathematical simplicity and physical plausibility. Lower values of $n$ may produce unphysical mass distributions, whereas higher values entail substantially greater computational cost due to steeper density gradients. Notably, in both models analysed here, the index $n_x$ (associated with the major axis, typically considered the defining direction of the bar) remains close to $n=2$ throughout the evolution. This lends further support to the continued use of $n=2$ in theoretical studies as a reasonable approximation, despite its limitations.

However, the fact that $n_x$ remains close to $n=2$ along the major axis does not imply that a constant and isotropic Ferrers index fully captures the structural evolution of the bar. Our results show that $n_x$, $n_y$, and $n_z$ evolve differently over time, revealing that the bar develops anisotropic density profiles that cannot be described by a single constant index. This highlights an important limitation of standard Ferrers models, which assume $n_x = n_y = n_z$ and therefore neglect directional variations in the bar structure.

The methodology presented here provides additional insight by quantifying how the density profile evolves independently along each principal axis. This allows us to identify phases in which the bar becomes more centrally concentrated or more extended in specific directions, reflecting the underlying dynamical processes such as orbital trapping, secular evolution, and structural reorganization. Therefore, while a Ferrers model with $n=2$ remains a reasonable first-order approximation for the overall structure of the bar, our approach reveals important deviations from this idealized model and provides a more realistic description of the bar internal evolution in self-consistent simulations.

This demonstrates that self-consistent bars cannot, in general, be fully characterized by a single structural parameter, and motivates the development of more realistic analytical models that incorporate axis-dependent density profiles.

\subsection{Comparison between models}

Although models $A\lambda03$ and $A\lambda04$ both form bars, their dynamical evolution proceeds along markedly different pathways. Differences in the development of orbital chaos, the trapping efficiency of $x_1$ orbits, and the growth of the bar semi-axes indicate distinct regimes of bar assembly. In this subsection we summarize these contrasts using both spectral and structural diagnostics.

\subsubsection{Orbital structure and chaos}

The temporal evolution of the minimal spectral entropy $S_{\min}$ and the fraction of bar-supporting $x_1$ orbits, $P_{x_1}$ (Fig.~\ref{fig:Smax_vs_Px1}), shows that bar formation in both models is accompanied by an increase in chaoticity and a simultaneous reduction in $P_{x_1}$. However, the character of this chaotic phase differs substantially.

In $A\lambda03$, the fast bar formation comes with a rapid and intense rise in $S_{\min}$, indicating that a large fraction of orbits undergoes an abrupt transition toward chaotic motion. This phase coincides with sharp drops in $P_{x_1}$, implying significant destabilization of bar-supporting trajectories. Although $P_{x_1}$ later recovers, its overall level remains systematically below that of $A\lambda04$, suggesting less efficient long-term trapping into the $x_1$ family.

By contrast, $A\lambda04$ exhibits a milder but more prolonged increase in $S_{\min}$. Despite the extended presence of chaos, $P_{x_1}$ grows more steadily and ultimately reaches higher values than in $A\lambda03$. This behaviour indicates that chaos in $A\lambda04$ is less disruptive, likely dominated by sticky orbits that can eventually be incorporated into the bar-supporting family. Thus, while both models experience chaotic phases, their qualitative nature strongly influences the efficiency of orbital trapping and the persistence of the bar.

Additionally, during the first $\sim 1$ Gyr, before or during the earliest stages of bar formation, $A\lambda03$ progressively shifts toward higher $S_{\min}$ values, whereas $A\lambda04$ remains comparatively more stable, with strong fluctuations around a nearly flat trend. This suggests that the early increase of $S_{\min}$ in $A\lambda03$ is associated with the dynamical response of the system to its more unstable configuration. In this sense, the spectral entropy acts as a diagnostic of the orbital response related to the distinct dynamical pathways followed by the two models.

These differences can be interpreted in light of the disc spin parameter. According to \citet{2019AJ....157..175V}, bar formation is favoured when $\lambda_d < \lambda_c$, and the magnitude of the difference $\lambda_c - \lambda_d$ affects the bar growth timescale, with larger separations leading to more rapid growth. As shown in their analysis, this separation is larger for $A\lambda03$ than for $A\lambda04$. The stronger instability in $A\lambda03$ is therefore consistent with its faster bar formation and enhanced chaoticity, while the smaller separation in $A\lambda04$ leads to a more gradual evolution and weaker orbital disruption. Although this interpretation aligns with the theoretical framework of \citet{2019AJ....157..175V}, a direct quantitative link between $\lambda_c - \lambda_d$ and the measured spectral entropy would require further investigation.

\subsubsection{Structural evolution}

The distinct dynamical histories of the two models are also reflected in the evolution of the bar semi-axes (Fig.~\ref{fig:semiaxes}). 

Model $A\lambda03$ undergoes a rapid initial growth phase in which the major semi-axis $a_B$ increases sharply, followed by a reorganization phase characterized by mild oscillations and temporary contraction. During the subsequent secular phase, all three semi-axes grow steadily, with $a_B$ increasing more rapidly than $b_B$ and $c_B$, leading to a progressively more elongated structure.

In contrast, $A\lambda04$ develops a significantly longer bar during its early evolution, with $a_B$ reaching substantially larger values. Strong oscillations in $a_B$ are present throughout the growth and secular phases, while $b_B$ and $c_B$ evolve more smoothly. Unlike $A\lambda03$, no clear contraction stage is observed. The persistent oscillatory behaviour of $a_B$ suggests repeated episodes of particle trapping and release near the bar ends. Indeed, variations in $a_B$ correlate with changes in $P_{x_1}$, indicating a coupling between bar length and the efficiency of orbital capture. However, the large oscillations during the earliest stages of $A\lambda04$ are not entirely mirrored by $P_{x_1}$, implying that additional dynamical processes contribute to the modulation of the bar length.

Further insight is provided by the evolution of the structural indices $n_x$, $n_y$, and $n_z$ (Fig.~\ref{fig:index_ev}). In both models, $n_x$ remains close to $n=2$, consistent with a relatively shallow density profile along the major axis. In contrast, $n_y$ and $n_z$ exceed $n=2$ for most of the evolution, reflecting increasing concentration perpendicular to the bar. 

Model $A\lambda03$ exhibits abrupt variations in these indices near the end of its reorganization phase, followed by smoother secular growth, consistent with an early episode of strong mass redistribution. Model $A\lambda04$ shows similar transitional features but with more persistent fluctuations in $n_y$, indicative of continued structural adjustments. Despite these oscillations, its consistently elevated perpendicular indices and sustained growth of $a_B$ point to a stronger and more resilient bar configuration.

\subsubsection{Overall comparison}

Taken together, the spectral and structural diagnostics consistently indicate that $A\lambda03$ forms its bar under a stronger dynamical instability, leading to rapid growth, intense chaotic restructuring, and less efficient long-term trapping of bar-supporting orbits. In contrast, $A\lambda04$ evolves through a milder but more sustained dynamical regime that favours gradual orbital organization, higher $x_1$ occupation, and the development of a longer and more persistent bar. These results highlight how differences in the spin parameter regime translate into distinct pathways of bar assembly and structural stabilization.

\section{Discussions and Conclusions}
\label{sec:conclusions}

In this paper, we introduced a new method to identify sticky orbits associated with two of the principal orbital families in barred galaxies: the $x_1$ family, which primarily supports the bar structure, and the $x_2$ family, typically related to a secondary bar component oriented perpendicular to the main bar. Our approach enables the systematic identification of particles trapped in the vicinity of these families, particularly those exhibiting sticky behaviour around the $x_1$ periodic orbits.

By isolating sticky orbits associated with the $x_1$ family, we demonstrate that our method not only reliably recovers the dynamical backbone of the bar, but also provides additional insight into the extent and structure of the bar-supported region. These results validate the robustness of our approach and highlight its potential as a diagnostic tool for characterizing bar dynamics in $N$-body simulations. Below, we summarize our main findings.

\subsection{Main Findings}

\subsubsection{Validation against established bar diagnostics}

We validated our method by comparing the radial Fourier amplitude $A_2(R,t)$ with both $P_{x_1}(t)$ and the bar semi-major axis $a_B(t)$ (Fig. \ref{fig:A2_vs_Px1}). We find that $P_{x_1}$ closely follows the amplitude of $A_2$ within the bar region, indicating that our identification of sticky $x_1$ orbits captures the same dynamical component responsible for the bar strength. Moreover, the semi-major axis $a_B$ derived from our method agrees with the radial extent over which $A_2$ reaches its maximum, supporting the consistency of our approach in determining the bar size.

We further compared the frozen-potential methodology with different bar diagnostics in Fig. \ref{fig:comp_method}. Among the different approaches considered, the temporal evolution of $P_{x_1}$ derived from our method is the one that most closely reproduces the behaviour of the global Fourier amplitude $\bar{A}_2$, one of the standard measures of bar strength in the literature. This agreement indicates that our orbital-based quantification captures the same global trends traditionally inferred from Fourier analysis. However, unlike $\bar{A}_2$, our method provides a direct physical interpretation by explicitly identifying the particles trapped around specific orbital families. In addition, it naturally extends to three dimensions, allowing us to characterize the full spatial structure of the bar, an aspect that purely Fourier-based diagnostics cannot address.

Although the frozen-potential approach has intrinsic limitations, it offers significant advantages. Most importantly, it enables a robust classification of particles into distinct orbital families and provides detailed information about their dynamical and structural properties. This capability makes it a powerful tool for dissecting the internal orbital composition of barred galaxies.

\subsubsection{Angular momentum redistribution}
An important result arises from comparing the bar semi-major axis with the angular momentum distribution. We find that the bar region corresponds to a low-angular-momentum zone in the disc. After its formation, the bar becomes the main driver of angular momentum redistribution, promoting its transfer from the inner to the outer regions.

In our models, the bar plays a dual role: it acts as a sink of angular momentum in the inner disc, where particles trapped in bar-supporting orbits lose angular momentum, while simultaneously driving its outward transport. This process sustains bar growth and shapes the secular evolution of the galaxy, reinforcing the view of the bar as a dynamical engine rather than merely a morphological feature.

These results are consistent with the theoretical framework of \citet{Athanassoula2003, 2003MNRAS.341.1179A, Athanassoula2013}, in which bar evolution is governed by angular momentum exchange between galactic components. They also agree with \citet{10.1093mnrasstz2824}, who identified the bar’s dual role through direct torque measurements. The agreement between our findings and previous studies strengthens the robustness of our analysis and supports the reliability of our orbital-based method in capturing the fundamental mechanisms driving bar evolution.

\subsubsection{The role of chaos}

We further explored the connection between bar evolution and the global dynamical state of the disc by relating the bar mass fraction at different evolutionary stages (i. The chaotic birth phase, ii. The structural reorganization phase, and iii. The long-term secular stage) to the distribution and temporal evolution of $S_{\min}$.

We find a clear correlation between the bar mass fraction and the overall degree of orbital chaoticity, although this relation varies across evolutionary stages. Either weakening or strengthening the bar depending on the prevailing level of chaoticity. These results suggest that orbital chaos is not merely a consequence of bar evolution, but a key factor regulating its strength and long-term development through its interplay with regular bar-supporting orbits.

\subsubsection{Bar shape and density profile}

By tracking the evolution of the bar semi-axes and the steepness of its density profile along each principal direction, we find that both quantities evolve at different rates along the three axes. This indicates that the bar does not simply increase in size, but undergoes continuous changes in shape and internal mass distribution.

An important result is that the density profile indices satisfy $n_x \neq n_y \neq n_z$ during most of the simulation time. This anisotropic behaviour contrasts with the common assumption adopted in many analytical models, where a single, constant index is used ($n = n_x = n_y = n_z$), as in the classical Ferrers profile. Our results show that such an assumption represents a significant simplification and does not capture the evolving, non-axisymmetric nature of self-consistent bars in $N$-body simulations.

Moreover, the indices along the three axes are rarely equal to the canonical value $n=2$ commonly adopted for Ferrers bars. This suggests that the widespread use of $n=2$ is largely motivated by mathematical convenience rather than by direct dynamical evidence. Although $n_x$ approaches 2 during the late secular stage (possibly explaining why this approximation has been practically useful) our results show that assuming $n_x = n_y = n_z = 2$ does not reproduce the anisotropic and time-dependent structure observed in self-consistent $N$-body simulations such as those analysed here.

\subsubsection{Bar and pseudo-bulge connection}

Our method allows us to reliably identify bar particles and separate them from those belonging to other galactic components. This distinction is crucial, as it prevents bar particles from being misclassified as part of the bulge or pseudo-bulge, an issue that arises when applying the methodology of \citet{2022MNRAS.515.1524Z} without explicitly identifying the bar. As a result, we are able to perform a more meaningful comparison of the evolutionary paths of these components.

A critical result is the apparent anti-correlation between the bar and pseudo-bulge fractions, suggesting an ongoing exchange of stars between the pseudo-bulge and the bar, with the pseudo-bulge representing one of the main stellar reservoirs contributing to bar growth.

Overall, our results demonstrate that accounting for the bar is essential for obtaining a physically meaningful decomposition of barred galaxies and for accurately tracing the secular evolution of their stellar components.

\subsubsection{Dynamical differences between models}

Comparing the two models, we find that although $A\lambda03$ forms a bar more rapidly, this process is accompanied by a higher level of orbital chaoticity. The strong and abrupt perturbations associated with this rapid growth hinder the efficient trapping of particles into bar-supporting families, resulting in a weaker and less stable bar.

In contrast, $A\lambda04$ evolves more gradually, allowing stellar orbits to reorganize as the bar potential develops. This smoother evolution favours the long-term trapping of particles into bar-supporting orbits, leading to a stronger, more extended, and more stable bar. These results indicate that the timescale of bar formation plays an important role in regulating both chaoticity and the structural robustness of the bar.

This behaviour is consistent with the spin parameter criterion of \citet{2019AJ....157..175V}. In this framework, the larger separation between $\lambda_d$ and $\lambda_c$ in $A\lambda03$ leads to a stronger and more violent instability, producing higher chaoticity and less efficient orbital trapping. Conversely, the smaller separation in $A\lambda04$ results in a milder instability, promoting gradual orbital organization and the formation of a more stable bar.

\subsection{Final conclusions}

The results presented in this work  validate our approach through its consistency with established results in the literature and demonstrate its capability to uncover new aspects of bar dynamics. In particular, our analysis shows that bar evolution is governed by a complex interplay between orbital structure, angular momentum redistribution, and the degree of chaoticity in the disc.

A strength of our method is its ability to directly identify the orbital families that support the bar and to track their contribution over time. This provides a physically grounded description of bar evolution that goes beyond traditional diagnostics. In this context, our framework offers a promising basis for future studies aimed at connecting theoretical models with observations. For instance, extending this approach to simulations including gas could enable the generation of mock observables and the exploration of links between bar properties and spectral or kinematic signatures.

At present, our analysis is limited to isolated galaxy simulations, and its applicability to interacting systems or fully cosmological environments remains to be tested. In such cases, additional challenges particularly in accurately computing the gravitational potential, must be addressed. Nevertheless, we expect that the methodology can be extended to these more complex scenarios.

In conclusion, the method introduced here provides a powerful tool for identifying bar-supporting orbital families and quantifying their structural and dynamical evolution. Despite the computational cost associated with the frozen-potential approach, its ability to robustly separate orbital families and recover their properties makes it especially valuable for advancing our understanding of barred galaxy dynamics.


\section*{Acknowledgements}

The authors thank the anonymous referee for his/her careful reading of the manuscript and for constructive comments and suggestions which significantly improved the quality and clarity of this work.
A.S.C. would like to thank the Secretaría de Ciencia, Humanidades, Tecnología e Innovación (SECIHTI) for funding a scholarship, and the Instituto Nacional de Astrofísica, Óptica y Electrónica (INAOE) for providing the facilities and support that made this work possible.
I.P. thanks the former Mexican Foundation CONACYT for funds used to the acquisition of the cluster Olinki in which all the calculations for this study were conducted. 
D.V.E. acknowledges support from the {Facultad de Ingenier\'{\i}a} and {Direcci\'{o}n de Investigaciones} of Universidad Mariana (projects IC1-17 and INTD2-61).

\section*{Data Availability}

The data underlying this article will be shared on reasonable request
to the corresponding author.



\bibliographystyle{mnras}
\bibliography{example} 

@ARTICLE{2012ApJ...757...60K,
       author = {{Kraljic}, Katarina and {Bournaud}, Fr{\'e}d{\'e}ric and {Martig}, Marie},
        title = "{The Two-phase Formation History of Spiral Galaxies Traced by the Cosmic Evolution of the Bar Fraction}",
      journal = {\apj},
         year = 2012,
        month = sep,
       volume = {757},
       number = {1},
          eid = {60},
        pages = {60},
          doi = {10.1088/0004-637X/757/1/60},
archivePrefix = {arXiv},
       eprint = {1207.0351},
 primaryClass = {astro-ph.GA},
       adsurl = {https://ui.adsabs.harvard.edu/abs/2012ApJ...757...60K}
}

@ARTICLE{2014MNRAS.438.2882M,
       author = {{Melvin}, Thomas and {Masters}, Karen and {Lintott}, Chris and {Nichol}, Robert C. and {Simmons}, Brooke and {Bamford}, Steven P. and {Casteels}, Kevin R.~V. and {Cheung}, Edmond and {Edmondson}, Edward M. and {Fortson}, Lucy and {Schawinski}, Kevin and {Skibba}, Ramin A. and {Smith}, Arfon M. and {Willett}, Kyle W.},
        title = "{Galaxy Zoo: an independent look at the evolution of the bar fraction over the last eight billion years from HST-COSMOS}",
      journal = {\mnras},
         year = 2014,
        month = mar,
       volume = {438},
       number = {4},
        pages = {2882-2897},
          doi = {10.1093/mnras/stt2397},
archivePrefix = {arXiv},
       eprint = {1401.3334},
 primaryClass = {astro-ph.GA},
       adsurl = {https://ui.adsabs.harvard.edu/abs/2014MNRAS.438.2882M}
}

@ARTICLE{2025ApJ...987...74G,
       author = {{G{\'e}ron}, Tobias and {Smethurst}, R.~J. and {Dickinson}, Hugh and {Fortson}, L.~F. and {Garland}, Izzy L. and {Kruk}, Sandor and {Lintott}, Chris and {Makechemu}, Jason Shingirai and {Mantha}, Kameswara Bharadwaj and {Masters}, Karen L. and {O'Ryan}, David and {Roberts}, Hayley and {Simmons}, B.~D. and {Walmsley}, Mike and {Calabr{\`o}}, Antonello and {Chiba}, Rimpei and {Costantin}, Luca and {Drout}, Maria R. and {Fragkoudi}, Francesca and {Guo}, Yuchen and {Holwerda}, B.~W. and {Jogee}, Shardha and {Koekemoer}, Anton M. and {Lucas}, Ray A. and {Pacucci}, Fabio},
        title = "{Galaxy Zoo CEERS: Bar Fractions Up to z {\ensuremath{\sim}} 4.0}",
      journal = {\apj},
         year = 2025,
        month = jul,
       volume = {987},
       number = {1},
          eid = {74},
        pages = {74},
          doi = {10.3847/1538-4357/add7d0},
archivePrefix = {arXiv},
       eprint = {2505.01421},
 primaryClass = {astro-ph.GA},
       adsurl = {https://ui.adsabs.harvard.edu/abs/2025ApJ...987...74G}
}

@ARTICLE{2024MNRAS.530.1984L,
       author = {{Le Conte}, Zoe A. and {Gadotti}, Dimitri A. and {Ferreira}, Leonardo and {Conselice}, Christopher J. and {de S{\'a}-Freitas}, Camila and {Kim}, Taehyun and {Neumann}, Justus and {Fragkoudi}, Francesca and {Athanassoula}, E. and {Adams}, Nathan J.},
        title = "{A JWST investigation into the bar fraction at redshifts 1 {\ensuremath{\leq}} z {\ensuremath{\leq}} 3}",
      journal = {\mnras},
         year = 2024,
        month = may,
       volume = {530},
       number = {2},
        pages = {1984-2000},
          doi = {10.1093/mnras/stae921},
archivePrefix = {arXiv},
       eprint = {2309.10038},
 primaryClass = {astro-ph.GA},
       adsurl = {https://ui.adsabs.harvard.edu/abs/2024MNRAS.530.1984L}
}

@ARTICLE{2023ApJ...947...80B,
       author = {{Bland-Hawthorn}, Joss and {Tepper-Garcia}, Thor and {Agertz}, Oscar and {Freeman}, Ken},
        title = "{The Rapid Onset of Stellar Bars in the Baryon-dominated Centers of Disk Galaxies}",
      journal = {\apj},
         year = 2023,
        month = apr,
       volume = {947},
       number = {2},
          eid = {80},
        pages = {80},
          doi = {10.3847/1538-4357/acc469},
archivePrefix = {arXiv},
       eprint = {2303.05574},
 primaryClass = {astro-ph.GA},
       adsurl = {https://ui.adsabs.harvard.edu/abs/2023ApJ...947...80B}
}

@ARTICLE{2004ApJ...612..191E,
       author = {{Elmegreen}, Bruce G. and {Elmegreen}, Debra Meloy and {Hirst}, Amelia C.},
        title = "{A Constant Bar Fraction out to Redshift z \raisebox{-0.5ex}\textasciitilde 1 in the Advanced Camera for Surveys Field of the Tadpole Galaxy}",
      journal = {\apj},
         year = 2004,
        month = sep,
       volume = {612},
       number = {1},
        pages = {191-201},
          doi = {10.1086/422407},
archivePrefix = {arXiv},
       eprint = {astro-ph/0407577},
 primaryClass = {astro-ph},
       adsurl = {https://ui.adsabs.harvard.edu/abs/2004ApJ...612..191E}
}

@ARTICLE{2025MNRAS.542..151M,
       author = {{Mukundan}, Kavya and {Nair}, Preethi and {Masters}, Karen L. and {Bailin}, Jeremy and {Gwartney}, Peter and {Li}, Wenhao},
        title = "{Bar fraction and its dependence on host galaxy properties in the local Universe}",
      journal = {\mnras},
         year = 2025,
        month = sep,
       volume = {542},
       number = {1},
        pages = {151-169},
          doi = {10.1093/mnras/staf1143},
       adsurl = {https://ui.adsabs.harvard.edu/abs/2025MNRAS.542..151M}
}

@ARTICLE{2015ApJS..217...32B,
       author = {{Buta}, Ronald J. and {Sheth}, Kartik and {Athanassoula}, E. and {Bosma}, A. and {Knapen}, Johan H. and {Laurikainen}, Eija and {Salo}, Heikki and {Elmegreen}, Debra and {Ho}, Luis C. and {Zaritsky}, Dennis and {Courtois}, Helene and {Hinz}, Joannah L. and {Mu{\~n}oz-Mateos}, Juan-Carlos and {Kim}, Taehyun and {Regan}, Michael W. and {Gadotti}, Dimitri A. and {Gil de Paz}, Armando and {Laine}, Jarkko and {Men{\'e}ndez-Delmestre}, Kar{\'\i}n and {Comer{\'o}n}, S{\'e}bastien and {Erroz Ferrer}, Santiago and {Seibert}, Mark and {Mizusawa}, Trisha and {Holwerda}, Benne and {Madore}, Barry F.},
        title = "{A Classical Morphological Analysis of Galaxies in the Spitzer Survey of Stellar Structure in Galaxies (S4G)}",
      journal = {\apjs},
         year = 2015,
        month = apr,
       volume = {217},
       number = {2},
          eid = {32},
        pages = {32},
          doi = {10.1088/0067-0049/217/2/32},
archivePrefix = {arXiv},
       eprint = {1501.00454},
 primaryClass = {astro-ph.GA},
       adsurl = {https://ui.adsabs.harvard.edu/abs/2015ApJS..217...32B}
}

@ARTICLE{2002AJ....124...65E,
       author = {{Erwin}, Peter and {Sparke}, Linda S.},
        title = "{Double Bars, Inner Disks, and Nuclear Rings in Early-Type Disk Galaxies}",
      journal = {\aj},
         year = 2002,
        month = jul,
       volume = {124},
       number = {1},
        pages = {65-77},
          doi = {10.1086/340803},
archivePrefix = {arXiv},
       eprint = {astro-ph/0203514},
 primaryClass = {astro-ph},
       adsurl = {https://ui.adsabs.harvard.edu/abs/2002AJ....124...65E}
}

@ARTICLE{1993A&A...277...27F,
       author = {{Friedli}, D. and {Martinet}, L.},
        title = "{Bars within bars in lenticular and spiral galaxies : a step in secular evolution?}",
      journal = {\aap},
         year = 1993,
        month = sep,
       volume = {277},
        pages = {27-41},
       adsurl = {https://ui.adsabs.harvard.edu/abs/1993A&A...277...27F}
}

@article{Martinez_Valpuesta_2006,
doi = {10.1086/498338},
url = {https://doi.org/10.1086/498338},
year = {2006},
month = {jan},
publisher = {},
volume = {637},
number = {1},
pages = {214},
author = {Martinez-Valpuesta, Inma and Shlosman, Isaac and Heller, Clayton},
title = {Evolution of Stellar Bars in Live Axisymmetric Halos: Recurrent Buckling and Secular Growth},
journal = {The Astrophysical Journal}
}

@article{sellwood2014,
  title = {Secular evolution in disk galaxies},
  author = {Sellwood, J. A.},
  journal = {Rev. Mod. Phys.},
  volume = {86},
  issue = {1},
  pages = {1--46},
  numpages = {0},
  year = {2014},
  month = {Jan},
  publisher = {American Physical Society},
  doi = {10.1103/RevModPhys.86.1},
  url = {https://link.aps.org/doi/10.1103/RevModPhys.86.1}
}

@article{skokos2002,
    author = {Skokos, Ch. and Patsis, P. A. and Athanassoula, E.},
    title = {Orbital dynamics of three-dimensional bars – I. The backbone of three-dimensional bars. A fiducial case},
    journal = {Monthly Notices of the Royal Astronomical Society},
    volume = {333},
    number = {4},
    pages = {847-860},
    year = {2002},
    month = {07},
    issn = {0035-8711},
    doi = {10.1046/j.1365-8711.2002.05468.x},
    url = {https://doi.org/10.1046/j.1365-8711.2002.05468.x},
    eprint = {https://academic.oup.com/mnras/article-pdf/333/4/847/3090229/333-4-847.pdf},
}

@article{athanassoula1992,
    author = {Athanassoula, E.},
    title = {Morphology of bar orbits},
    journal = {Monthly Notices of the Royal Astronomical Society},
    volume = {259},
    number = {2},
    pages = {328-344},
    year = {1992},
    month = {11},
    issn = {0035-8711},
    doi = {10.1093/mnras/259.2.328},
    url = {https://doi.org/10.1093/mnras/259.2.328},
    eprint = {https://academic.oup.com/mnras/article-pdf/259/2/328/3618901/mnras259-0328.pdf},
}

@article{contopoulos1980orbits,
  title={Orbits in weak and strong bars},
  author={Contopoulos, George and Papayannopoulos, Th},
  journal={Astronomy and Astrophysics, vol. 92, no. 1-2, Dec. 1980, p. 33-46.},
  volume={92},
  pages={33--46},
  year={1980}
}

@article{Trapp2024,
    author = {Trapp, Cameron W and Kereš, Dušan and Hopkins, Philip F and Faucher-Giguère, Claude-André and Murray, Norman},
    title = {Angular momentum transfer in cosmological simulations of Milky Way-mass discs},
    journal = {Monthly Notices of the Royal Astronomical Society},
    volume = {533},
    number = {3},
    pages = {3008-3026},
    year = {2024},
    month = {08},
    issn = {0035-8711},
    doi = {10.1093/mnras/stae2021},
    url = {https://doi.org/10.1093/mnras/stae2021},
    eprint = {https://academic.oup.com/mnras/article-pdf/533/3/3008/58976110/stae2021.pdf},
}

@Inbook{Athanassoula2003,
author="Athanassoula, Lia",
editor="Contopoulos, George
and Voglis, Nikos",
title="Angular Momentum Redistribution and the Evolution and Morphology of Bars",
bookTitle="Galaxies and Chaos",
year="2003",
publisher="Springer Berlin Heidelberg",
address="Berlin, Heidelberg",
pages="313--326",
isbn="978-3-540-45040-5",
doi="10.1007/978-3-540-45040-5_26",
url="https://doi.org/10.1007/978-3-540-45040-5_26"
}

@ARTICLE{2003MNRAS.341.1179A,
       author = {{Athanassoula}, E.},
        title = "{What determines the strength and the slowdown rate of bars?}",
      journal = {\mnras},
         year = 2003,
        month = jun,
       volume = {341},
       number = {4},
        pages = {1179-1198},
          doi = {10.1046/j.1365-8711.2003.06473.x},
archivePrefix = {arXiv},
       eprint = {astro-ph/0302519},
 primaryClass = {astro-ph},
       adsurl = {https://ui.adsabs.harvard.edu/abs/2003MNRAS.341.1179A}
}

@incollection{Athanassoula2013,
  author       = {Athanassoula, E.},
  title        = {Bars and secular evolution in disk galaxies: Theoretical input},
  booktitle    = {Secular Evolution of Galaxies},
  editor       = {Falc{\'o}n-Barroso, J. and Knapen, J. H.},
  publisher    = {Cambridge University Press},
  address      = {Cambridge, UK},
  year         = {2013},
  pages        = {305--352},
  doi          = {10.48550/arXiv.1211.6752},
  archivePrefix = {arXiv},
  eprint       = {1211.6752},
  primaryClass = {astro-ph.GA},
  note         = {Proceedings of the XXIII Canary Islands Winter School of Astrophysics},
  bibcode      = {2013seg..book..305A}
}

@article{10.1093mnrasstz2824,
    author = {Petersen, Michael S and Weinberg, Martin D and Katz, Neal},
    title = {Using torque to understand barred galaxy models},
    journal = {Monthly Notices of the Royal Astronomical Society},
    volume = {490},
    number = {3},
    pages = {3616-3632},
    year = {2019},
    month = {10},
    issn = {0035-8711},
    doi = {10.1093/mnras/stz2824},
    url = {https://doi.org/10.1093/mnras/stz2824},
    eprint = {https://academic.oup.com/mnras/article-pdf/490/3/3616/30338159/stz2824.pdf},
}

@ARTICLE{2019AJ....157..175V,
       author = {{Valencia-Enr{\'\i}quez}, D. and {Puerari}, I. and {Rodrigues}, I.},
        title = "{Assessing Disk Galaxy Stability through Time}",
      journal = {\aj},
         year = 2019,
        month = may,
       volume = {157},
       number = {5},
          eid = {175},
        pages = {175},
          doi = {10.3847/1538-3881/ab100f},
archivePrefix = {arXiv},
       eprint = {1903.07728},
 primaryClass = {astro-ph.GA},
       adsurl = {https://ui.adsabs.harvard.edu/abs/2019AJ....157..175V}
}

@ARTICLE{2023MNRAS.525.3162V,
       author = {{Valencia-Enr{\'\i}quez}, Diego and {Puerari}, Iv{\^a}nio and {Chaves-Velasquez}, Leonardo},
        title = "{Orbital structure evolution in self-consistent N-body simulations}",
      journal = {\mnras},
         year = 2023,
        month = oct,
       volume = {525},
       number = {2},
        pages = {3162-3180},
          doi = {10.1093/mnras/stad2437},
archivePrefix = {arXiv},
       eprint = {2308.01439},
 primaryClass = {astro-ph.GA},
       adsurl = {https://ui.adsabs.harvard.edu/abs/2023MNRAS.525.3162V}
}

@ARTICLE{1999MNRAS.307..162S,
       author = {{Springel}, Volker and {White}, Simon D.~M.},
        title = "{Tidal tails in cold dark matter cosmologies}",
      journal = {\mnras},
         year = 1999,
        month = jul,
       volume = {307},
       number = {1},
        pages = {162-178},
          doi = {10.1046/j.1365-8711.1999.02613.x},
archivePrefix = {arXiv},
       eprint = {astro-ph/9807320},
 primaryClass = {astro-ph},
       adsurl = {https://ui.adsabs.harvard.edu/abs/1999MNRAS.307..162S}
}

@ARTICLE{2001NewA....6...79S,
       author = {{Springel}, Volker and {Yoshida}, Naoki and {White}, Simon D.~M.},
        title = "{GADGET: a code for collisionless and gasdynamical cosmological simulations}",
      journal = {\na},
         year = 2001,
        month = apr,
       volume = {6},
       number = {2},
        pages = {79-117},
          doi = {10.1016/S1384-1076(01)00042-2},
archivePrefix = {arXiv},
       eprint = {astro-ph/0003162},
 primaryClass = {astro-ph},
       adsurl = {https://ui.adsabs.harvard.edu/abs/2001NewA....6...79S}
}

@ARTICLE{2005MNRAS.364.1105S,
       author = {{Springel}, Volker},
        title = "{The cosmological simulation code GADGET-2}",
      journal = {\mnras},
         year = 2005,
        month = dec,
       volume = {364},
       number = {4},
        pages = {1105-1134},
          doi = {10.1111/j.1365-2966.2005.09655.x},
archivePrefix = {arXiv},
       eprint = {astro-ph/0505010},
 primaryClass = {astro-ph},
       adsurl = {https://ui.adsabs.harvard.edu/abs/2005MNRAS.364.1105S}
}

@ARTICLE{2002ApJ...569L..83A,
       author = {{Athanassoula}, E.},
        title = "{Bar-Halo Interaction and Bar Growth}",
      journal = {\apjl},
         year = 2002,
        month = apr,
       volume = {569},
       number = {2},
        pages = {L83-L86},
          doi = {10.1086/340784},
archivePrefix = {arXiv},
       eprint = {astro-ph/0203368},
 primaryClass = {astro-ph},
       adsurl = {https://ui.adsabs.harvard.edu/abs/2002ApJ...569L..83A}
}

@ARTICLE{1996ApJ...462..563N,
       author = {{Navarro}, Julio F. and {Frenk}, Carlos S. and {White}, Simon D.~M.},
        title = "{The Structure of Cold Dark Matter Halos}",
      journal = {\apj},
         year = 1996,
        month = may,
       volume = {462},
        pages = {563},
          doi = {10.1086/177173},
archivePrefix = {arXiv},
       eprint = {astro-ph/9508025},
 primaryClass = {astro-ph},
       adsurl = {https://ui.adsabs.harvard.edu/abs/1996ApJ...462..563N}
}

@ARTICLE{1997ApJ...490..493N,
       author = {{Navarro}, Julio F. and {Frenk}, Carlos S. and {White}, Simon D.~M.},
        title = "{A Universal Density Profile from Hierarchical Clustering}",
      journal = {\apj},
         year = 1997,
        month = dec,
       volume = {490},
       number = {2},
        pages = {493-508},
          doi = {10.1086/304888},
archivePrefix = {arXiv},
       eprint = {astro-ph/9611107},
 primaryClass = {astro-ph},
       adsurl = {https://ui.adsabs.harvard.edu/abs/1997ApJ...490..493N}
}

@ARTICLE{2024ApJ...965...77C,
       author = {{Chantavat}, T. and {Yuma}, S. and {Malelohit}, P. and {Worrakitpoonpon}, T.},
        title = "{Morphological Evolution of Disk Galaxies and Their Concentration, Asymmetry, and Clumpiness (CAS) Properties in Simulations across Toomre's Q Parameter}",
      journal = {\apj},
         year = 2024,
        month = apr,
       volume = {965},
       number = {1},
          eid = {77},
        pages = {77},
          doi = {10.3847/1538-4357/ad3218},
archivePrefix = {arXiv},
       eprint = {2403.05003},
 primaryClass = {astro-ph.GA},
       adsurl = {https://ui.adsabs.harvard.edu/abs/2024ApJ...965...77C}
}

@ARTICLE{2022MNRAS.517.5660G,
       author = {{Garma-Oehmichen}, L. and {Hern{\'a}ndez-Toledo}, H. and {Aquino-Ort{\'\i}z}, E. and {Martinez-Medina}, L. and {Puerari}, I. and {Cano-D{\'\i}az}, M. and {Valenzuela}, O. and {V{\'a}zquez-Mata}, J.~A. and {G{\'e}ron}, T. and {Mart{\'\i}nez-V{\'a}zquez}, L.~A. and {Lane}, R.},
        title = "{SDSS IV MaNGA: bar pattern speed in Milky Way analogue galaxies}",
      journal = {\mnras},
         year = 2022,
        month = dec,
       volume = {517},
       number = {4},
        pages = {5660-5677},
          doi = {10.1093/mnras/stac3069},
archivePrefix = {arXiv},
       eprint = {2210.11424},
 primaryClass = {astro-ph.GA},
       adsurl = {https://ui.adsabs.harvard.edu/abs/2022MNRAS.517.5660G}
}

@ARTICLE{2025RMxAA..61...99S,
       author = {{Silva-Castro}, A. and {Puerari}, I.},
        title = "{Ferrers Bar Response Models: A Grid Calculation for Galactic Models}",
      journal = {\rmxaa},
         year = 2025,
        month = apr,
       volume = {61},
        pages = {99-110},
          doi = {10.48550/arXiv.2502.03344},
archivePrefix = {arXiv},
       eprint = {2502.03344},
 primaryClass = {astro-ph.GA},
       adsurl = {https://ui.adsabs.harvard.edu/abs/2025RMxAA..61...99S}
}

@ARTICLE{2018arXiv180208255V,
       author = {{Vasiliev}, Eugene},
        title = "{Agama reference documentation}",
      journal = {arXiv e-prints},
         year = 2018,
        month = feb,
          eid = {arXiv:1802.08255},
        pages = {arXiv:1802.08255},
          doi = {10.48550/arXiv.1802.08255},
archivePrefix = {arXiv},
       eprint = {1802.08255},
 primaryClass = {astro-ph.IM},
       adsurl = {https://ui.adsabs.harvard.edu/abs/2018arXiv180208255V}
}

@misc{2018asclsoft05008V,
       author = {{Vasiliev}, Eugene},
        title = "{AGAMA: Action-based galaxy modeling framework}",
 howpublished = {Astrophysics Source Code Library, record ascl:1805.008},
         year = 2018,
        month = may,
          eid = {ascl:1805.008},
       adsurl = {https://ui.adsabs.harvard.edu/abs/2018ascl.soft05008V}
}

@ARTICLE{2002MNRAS.333..861S,
       author = {{Skokos}, Ch. and {Patsis}, P.~A. and {Athanassoula}, E.},
        title = "{Orbital dynamics of three-dimensional bars - II. Investigation of the parameter space}",
      journal = {\mnras},
         year = 2002,
        month = jul,
       volume = {333},
       number = {4},
        pages = {861-870},
          doi = {10.1046/j.1365-8711.2002.05469.x},
archivePrefix = {arXiv},
       eprint = {astro-ph/0204078},
 primaryClass = {astro-ph},
       adsurl = {https://ui.adsabs.harvard.edu/abs/2002MNRAS.333..861S}
}

@ARTICLE{1993CeMDA..56..191L,
       author = {{Laskar}, Jacques},
        title = "{Frequency Analysis of a Dynamical System}",
      journal = {Celestial Mechanics and Dynamical Astronomy},
         year = 1993,
        month = mar,
       volume = {56},
       number = {1-2},
        pages = {191-196},
          doi = {10.1007/BF00699731},
       adsurl = {https://ui.adsabs.harvard.edu/abs/1993CeMDA..56..191L}
}

@BOOK{2008gady.book.....B,
       author = {{Binney}, James and {Tremaine}, Scott},
        title = "{Galactic Dynamics: Second Edition}",
        year = 2008,
        publisher={Princeton university press},
       adsurl = {https://ui.adsabs.harvard.edu/abs/2008gady.book.....B}
}

@book{information_theory,

author = {Cover, Thomas M and Thomas, Joy A},
publisher = {John Wiley \& Sons, Ltd},
title = {Elements of Information Theory},
chapter = {2},
pages = {13-55},
doi = {https://doi.org/10.1002/047174882X.ch2},
url = {https://onlinelibrary.wiley.com/doi/abs/10.1002/047174882X.ch2},
year = {2005}
}

@ARTICLE{1877QJPAM..14....1F,
       author = {{Ferrers}, N.~M.},
        title = "{On the Potentials, Ellipsoids, Ellipsoidal Shells, Elliptic Laminae, and Elliptics Rings, of Variable Densities}",
      journal = {The Quarterly Journal of Pure and Applied Mathematics},
         year = 1877,
        month = jan,
       volume = {14},
        pages = {1-22},
       adsurl = {https://ui.adsabs.harvard.edu/abs/1877QJPAM..14....1F}
}

@ARTICLE{1984A&A...134..373P,
       author = {{Pfenniger}, D.},
        title = "{The 3D dynamics of barred galaxies}",
      journal = {\aap},
         year = 1984,
        month = may,
       volume = {134},
       number = {2},
        pages = {373-386},
       adsurl = {https://ui.adsabs.harvard.edu/abs/1984A&A...134..373P}
}

@ARTICLE{2022MNRAS.515.1524Z,
       author = {{Zana}, Tommaso and {Lupi}, Alessandro and {Bonetti}, Matteo and {Dotti}, Massimo and {Rosas-Guevara}, Yetli and {Izquierdo-Villalba}, David and {Bonoli}, Silvia and {Hernquist}, Lars and {Nelson}, Dylan},
        title = "{Morphological decomposition of TNG50 galaxies: methodology and catalogue}",
      journal = {\mnras},
         year = 2022,
        month = sep,
       volume = {515},
       number = {1},
        pages = {1524-1543},
          doi = {10.1093/mnras/stac1708},
archivePrefix = {arXiv},
       eprint = {2206.04693},
 primaryClass = {astro-ph.GA},
       adsurl = {https://ui.adsabs.harvard.edu/abs/2022MNRAS.515.1524Z}
}

@ARTICLE{2023ApJ...953..173B,
       author = {{Beane}, Angus and {Hernquist}, Lars and {D'Onghia}, Elena and {Marinacci}, Federico and {Conroy}, Charlie and {Qi}, Jia and {Sales}, Laura V. and {Torrey}, Paul and {Vogelsberger}, Mark},
        title = "{Stellar Bars in Isolated Gas-rich Spiral Galaxies Do Not Slow Down}",
      journal = {\apj},
         year = 2023,
        month = aug,
       volume = {953},
       number = {2},
          eid = {173},
        pages = {173},
          doi = {10.3847/1538-4357/ace2b9},
archivePrefix = {arXiv},
       eprint = {2209.03364},
 primaryClass = {astro-ph.GA},
       adsurl = {https://ui.adsabs.harvard.edu/abs/2023ApJ...953..173B}
}

@ARTICLE{1998MNRAS.298....1C,
       author = {{Carpintero}, Daniel D. and {Aguilar}, Luis A.},
        title = "{Orbit classification in arbitrary 2D and 3D potentials}",
      journal = {\mnras},
         year = 1998,
        month = jul,
       volume = {298},
       number = {1},
        pages = {1-21},
          doi = {10.1046/j.1365-8711.1998.01320.x},
       adsurl = {https://ui.adsabs.harvard.edu/abs/1998MNRAS.298....1C}
}

@ARTICLE{2016ApJ...818..141V,
       author = {{Valluri}, Monica and {Shen}, Juntai and {Abbott}, Caleb and {Debattista}, Victor P.},
        title = "{A Unified Framework for the Orbital Structure of Bars and Triaxial Ellipsoids}",
      journal = {\apj},
         year = 2016,
        month = feb,
       volume = {818},
       number = {2},
          eid = {141},
        pages = {141},
          doi = {10.3847/0004-637X/818/2/141},
archivePrefix = {arXiv},
       eprint = {1512.03467},
 primaryClass = {astro-ph.GA},
       adsurl = {https://ui.adsabs.harvard.edu/abs/2016ApJ...818..141V}
}

@ARTICLE{1982ApJ...252..308B,
       author = {{Binney}, J. and {Spergel}, D.},
        title = "{Spectral stellar dynamics}",
      journal = {\apj},
         year = 1982,
        month = jan,
       volume = {252},
        pages = {308-321},
          doi = {10.1086/159559},
       adsurl = {https://ui.adsabs.harvard.edu/abs/1982ApJ...252..308B}
}

@ARTICLE{1984MNRAS.206..159B,
       author = {{Binney}, J. and {Spergel}, D.},
        title = "{Spectral stellar dynamics. II - The action integrals}",
      journal = {\mnras},
         year = 1984,
        month = jan,
       volume = {206},
        pages = {159-177},
          doi = {10.1093/mnras/206.1.159},
       adsurl = {https://ui.adsabs.harvard.edu/abs/1984MNRAS.206..159B}
}

@ARTICLE{2024MNRAS.531..751P,
       author = {{Petersen}, Michael S. and {Weinberg}, Martin D. and {Katz}, Neal},
        title = "{Measuring the dynamical length of galactic bars}",
      journal = {\mnras},
         year = 2024,
        month = jun,
       volume = {531},
       number = {1},
        pages = {751-763},
          doi = {10.1093/mnras/stae736},
archivePrefix = {arXiv},
       eprint = {2305.13366},
 primaryClass = {astro-ph.GA},
       adsurl = {https://ui.adsabs.harvard.edu/abs/2024MNRAS.531..751P}
}

@ARTICLE{2023ApJ...942..106J,
       author = {{Jang}, Dajeong and {Kim}, Woong-Tae},
        title = "{Effects of the Central Mass Concentration on Bar Formation in Disk Galaxies}",
      journal = {\apj},
         year = 2023,
        month = jan,
       volume = {942},
       number = {2},
          eid = {106},
        pages = {106},
          doi = {10.3847/1538-4357/aca7bc},
archivePrefix = {arXiv},
       eprint = {2211.16816},
 primaryClass = {astro-ph.GA},
       adsurl = {https://ui.adsabs.harvard.edu/abs/2023ApJ...942..106J}
}

@ARTICLE{2011MSAIS..18..145E,
       author = {{Erwin}, P.},
        title = "{Double-barred galaxies.}",
      journal = {Memorie della Societa Astronomica Italiana Supplementi},
         year = 2011,
        month = jan,
       volume = {18},
        pages = {145},
       adsurl = {https://ui.adsabs.harvard.edu/abs/2011MSAIS..18..145E}
}

@ARTICLE{2013ApJ...772...36G,
       author = {{Guedes}, Javiera and {Mayer}, Lucio and {Carollo}, Marcella and {Madau}, Piero},
        title = "{Pseudobulge Formation as a Dynamical Rather than a Secular Process}",
      journal = {\apj},
         year = 2013,
        month = jul,
       volume = {772},
       number = {1},
          eid = {36},
        pages = {36},
          doi = {10.1088/0004-637X/772/1/36},
archivePrefix = {arXiv},
       eprint = {1211.1713},
 primaryClass = {astro-ph.CO},
       adsurl = {https://ui.adsabs.harvard.edu/abs/2013ApJ...772...36G}
}

@ARTICLE{2011MNRAS.415.3308G,
       author = {{Gadotti}, Dimitri A.},
        title = "{Secular evolution and structural properties of stellar bars in galaxies}",
      journal = {\mnras},
         year = 2011,
        month = aug,
       volume = {415},
       number = {4},
        pages = {3308-3318},
          doi = {10.1111/j.1365-2966.2011.18945.x},
archivePrefix = {arXiv},
       eprint = {1003.1719},
 primaryClass = {astro-ph.CO},
       adsurl = {https://ui.adsabs.harvard.edu/abs/2011MNRAS.415.3308G}
}

@ARTICLE{2016MNRAS.463.1952P,
       author = {{Petersen}, Michael S. and {Weinberg}, Martin D. and {Katz}, Neal},
        title = "{Dark matter trapping by stellar bars: the shadow bar}",
      journal = {\mnras},
         year = 2016,
        month = dec,
       volume = {463},
       number = {2},
        pages = {1952-1967},
          doi = {10.1093/mnras/stw2141},
archivePrefix = {arXiv},
       eprint = {1602.04826},
 primaryClass = {astro-ph.GA},
       adsurl = {https://ui.adsabs.harvard.edu/abs/2016MNRAS.463.1952P}
}

@ARTICLE{2023MNRAS.518.2712D,
       author = {{Dehnen}, Walter and {Semczuk}, Marcin and {Sch{\"o}nrich}, Ralph},
        title = "{Measuring bar pattern speeds from single simulation snapshots}",
      journal = {\mnras},
         year = 2023,
        month = jan,
       volume = {518},
       number = {2},
        pages = {2712-2718},
          doi = {10.1093/mnras/stac3184},
archivePrefix = {arXiv},
       eprint = {2211.00674},
 primaryClass = {astro-ph.GA},
       adsurl = {https://ui.adsabs.harvard.edu/abs/2023MNRAS.518.2712D}
}

@ARTICLE{2025CNSNS.14908923S,
       author = {{S{\'a}nchez-Mart{\'\i}n}, P. and {Amor{\'o}s}, J. and {Masdemont}, J.~J.},
        title = "{Capturing the short-term characteristics of a barred galaxy from a single snapshot}",
      journal = {Communications in Nonlinear Science and Numerical Simulations},
         year = 2025,
        month = oct,
       volume = {149},
          eid = {108923},
        pages = {108923},
          doi = {10.1016/j.cnsns.2025.108923},
archivePrefix = {arXiv},
       eprint = {2502.02612},
 primaryClass = {astro-ph.GA},
       adsurl = {https://ui.adsabs.harvard.edu/abs/2025CNSNS.14908923S}
}

@ARTICLE{2016A&A...587A.160D,
       author = {{D{\'\i}az-Garc{\'\i}a}, S. and {Salo}, H. and {Laurikainen}, E. and {Herrera-Endoqui}, M.},
        title = "{Characterization of galactic bars from 3.6 {\ensuremath{\mu}}m S$^{4}$G imaging}",
      journal = {\aap},
         year = 2016,
        month = mar,
       volume = {587},
          eid = {A160},
        pages = {A160},
          doi = {10.1051/0004-6361/201526161},
archivePrefix = {arXiv},
       eprint = {1509.06743},
 primaryClass = {astro-ph.GA},
       adsurl = {https://ui.adsabs.harvard.edu/abs/2016A&A...587A.160D}
}

@ARTICLE{2005MNRAS.364..283E,
       author = {{Erwin}, Peter},
        title = "{How large are the bars in barred galaxies?}",
      journal = {\mnras},
         year = 2005,
        month = nov,
       volume = {364},
       number = {1},
        pages = {283-302},
          doi = {10.1111/j.1365-2966.2005.09560.x},
archivePrefix = {arXiv},
       eprint = {astro-ph/0508590},
 primaryClass = {astro-ph},
       adsurl = {https://ui.adsabs.harvard.edu/abs/2005MNRAS.364..283E}
}

@ARTICLE{2016A&A...595A..63V,
       author = {{Vera}, Matias and {Alonso}, Sol and {Coldwell}, Georgina},
        title = "{Effect of bars on the galaxy properties}",
      journal = {\aap},
         year = 2016,
        month = oct,
       volume = {595},
          eid = {A63},
        pages = {A63},
          doi = {10.1051/0004-6361/201628750},
archivePrefix = {arXiv},
       eprint = {1607.08643},
 primaryClass = {astro-ph.GA},
       adsurl = {https://ui.adsabs.harvard.edu/abs/2016A&A...595A..63V}
}

@ARTICLE{2023A&A...679A...5A,
       author = {{Aguerri}, J. Alfonso L. and {Cuomo}, Virginia and {Rojas-Roncero}, Azahara and {Morelli}, Lorenzo},
        title = "{Properties of barred galaxies with the environment. I. The case of the Virgo cluster}",
      journal = {\aap},
         year = 2023,
        month = nov,
       volume = {679},
          eid = {A5},
        pages = {A5},
          doi = {10.1051/0004-6361/202347500},
archivePrefix = {arXiv},
       eprint = {2309.11982},
 primaryClass = {astro-ph.GA},
       adsurl = {https://ui.adsabs.harvard.edu/abs/2023A&A...679A...5A}
}

@ARTICLE{2010ApJ...711L..61M,
       author = {{M{\'e}ndez-Abreu}, J. and {S{\'a}nchez-Janssen}, R. and {Aguerri}, J.~A.~L.},
        title = "{Which Galaxies Host Bars and Disks? A Study of the Coma Cluster}",
      journal = {\apjl},
         year = 2010,
        month = mar,
       volume = {711},
       number = {2},
        pages = {L61-L65},
          doi = {10.1088/2041-8205/711/2/L61},
archivePrefix = {arXiv},
       eprint = {1002.0583},
 primaryClass = {astro-ph.CO},
       adsurl = {https://ui.adsabs.harvard.edu/abs/2010ApJ...711L..61M}
}

@ARTICLE{2015A&A...576A.102A,
       author = {{Aguerri}, J.~A.~L. and {M{\'e}ndez-Abreu}, J. and {Falc{\'o}n-Barroso}, J. and {Amorin}, A. and {Barrera-Ballesteros}, J. and {Cid Fernandes}, R. and {Garc{\'\i}a-Benito}, R. and {Garc{\'\i}a-Lorenzo}, B. and {Gonz{\'a}lez Delgado}, R.~M. and {Husemann}, B. and {Kalinova}, V. and {Lyubenova}, M. and {Marino}, R.~A. and {M{\'a}rquez}, I. and {Mast}, D. and {P{\'e}rez}, E. and {S{\'a}nchez}, S.~F. and {van de Ven}, G. and {Walcher}, C.~J. and {Backsmann}, N. and {Cortijo-Ferrero}, C. and {Bland-Hawthorn}, J. and {del Olmo}, A. and {Iglesias-P{\'a}ramo}, J. and {P{\'e}rez}, I. and {S{\'a}nchez-Bl{\'a}zquez}, P. and {Wisotzki}, L. and {Ziegler}, B.},
        title = "{Bar pattern speeds in CALIFA galaxies. I. Fast bars across the Hubble sequence}",
      journal = {\aap},
         year = 2015,
        month = apr,
       volume = {576},
          eid = {A102},
        pages = {A102},
          doi = {10.1051/0004-6361/201423383},
archivePrefix = {arXiv},
       eprint = {1501.05498},
 primaryClass = {astro-ph.GA},
       adsurl = {https://ui.adsabs.harvard.edu/abs/2015A&A...576A.102A}
}

@ARTICLE{2023MNRAS.523.5823B,
       author = {{Bekki}, Kenji},
        title = "{A mechanism of bar formation in disc galaxies: Synchronization of apsidal precession}",
      journal = {\mnras},
         year = 2023,
        month = aug,
       volume = {523},
       number = {4},
        pages = {5823-5840},
          doi = {10.1093/mnras/stac3097},
archivePrefix = {arXiv},
       eprint = {2210.17132},
 primaryClass = {astro-ph.GA},
       adsurl = {https://ui.adsabs.harvard.edu/abs/2023MNRAS.523.5823B}
}

@ARTICLE{2025ApJ...979..166W,
       author = {{Worrakitpoonpon}, T.},
        title = "{Bar Instability and Formation Timescale across Toomre's Q Parameter and Central Mass Concentration: Slow Bar Formation or True Stability}",
      journal = {\apj},
         year = 2025,
        month = feb,
       volume = {979},
       number = {2},
          eid = {166},
        pages = {166},
          doi = {10.3847/1538-4357/ada35f},
archivePrefix = {arXiv},
       eprint = {2412.18098},
 primaryClass = {astro-ph.GA},
       adsurl = {https://ui.adsabs.harvard.edu/abs/2025ApJ...979..166W}
}

@ARTICLE{1990A&A...230...37G,
       author = {{Gerin}, M. and {Combes}, F. and {Athanassoula}, E.},
        title = "{The influence of galaxy interactions on stellar bars.}",
      journal = {\aap},
         year = 1990,
        month = apr,
       volume = {230},
        pages = {37-54},
       adsurl = {https://ui.adsabs.harvard.edu/abs/1990A&A...230...37G}
}

@ARTICLE{1991A&A...245L...5S,
       author = {{Sundin}, M. and {Sundelius}, B.},
        title = "{Unexpected behaviour in the rotation of perturbed barred galaxies}",
      journal = {\aap},
         year = 1991,
        month = may,
       volume = {245},
       number = {1},
        pages = {L5-L8},
       adsurl = {https://ui.adsabs.harvard.edu/abs/1991A&A...245L...5S}
}

@ARTICLE{1993A&A...280..105S,
       author = {{Sundin}, M. and {Donner}, K.~J. and {Sundelius}, B.},
        title = "{Change in angular velocity of perturbed galactic bars}",
      journal = {\aap},
         year = 1993,
        month = dec,
       volume = {280},
       number = {1},
        pages = {105-116},
       adsurl = {https://ui.adsabs.harvard.edu/abs/1993A&A...280..105S}
}

@ARTICLE{1991MNRAS.250..161L,
       author = {{Little}, Blane and {Carlberg}, R.~G.},
        title = "{The long-term evolution of barred galaxies}",
      journal = {\mnras},
         year = 1991,
        month = may,
       volume = {250},
        pages = {161-170},
          doi = {10.1093/mnras/250.1.161},
       adsurl = {https://ui.adsabs.harvard.edu/abs/1991MNRAS.250..161L}
}

@ARTICLE{2015MNRAS.454.3641F,
       author = {{Fanali}, R. and {Dotti}, M. and {Fiacconi}, D. and {Haardt}, F.},
        title = "{Bar formation as driver of gas inflows in isolated disc galaxies}",
      journal = {\mnras},
         year = 2015,
        month = dec,
       volume = {454},
       number = {4},
        pages = {3641-3652},
          doi = {10.1093/mnras/stv2247},
archivePrefix = {arXiv},
       eprint = {1509.08474},
 primaryClass = {astro-ph.GA},
       adsurl = {https://ui.adsabs.harvard.edu/abs/2015MNRAS.454.3641F}
}

@ARTICLE{2023ApJ...958...77L,
       author = {{Li}, Zhi and {Du}, Min and {Debattista}, Victor P. and {Shen}, Juntai and {Li}, Hui and {Liu}, Jie and {Vogelsberger}, Mark and {Beane}, Angus and {Marinacci}, Federico and {Sales}, Laura V.},
        title = "{How Nested Bars Enhance, Modulate, and Are Destroyed by Gas Inflows}",
      journal = {\apj},
         year = 2023,
        month = nov,
       volume = {958},
       number = {1},
          eid = {77},
        pages = {77},
          doi = {10.3847/1538-4357/acffb3},
archivePrefix = {arXiv},
       eprint = {2310.04666},
 primaryClass = {astro-ph.GA},
       adsurl = {https://ui.adsabs.harvard.edu/abs/2023ApJ...958...77L}
}

@ARTICLE{2025A&A...697A.236L,
       author = {{Lu}, Shuai and {Du}, Min and {Debattista}, Victor P.},
        title = "{IllustrisTNG insights: Factors affecting the presence of bars in disk galaxies}",
      journal = {\aap},
         year = 2025,
        month = may,
       volume = {697},
          eid = {A236},
        pages = {A236},
          doi = {10.1051/0004-6361/202453143},
archivePrefix = {arXiv},
       eprint = {2412.02255},
 primaryClass = {astro-ph.GA},
       adsurl = {https://ui.adsabs.harvard.edu/abs/2025A&A...697A.236L}
}

@ARTICLE{2024A&A...684A.179R,
       author = {{Rosas-Guevara}, Yetli and {Bonoli}, Silvia and {Misa Moreira}, Carmen and {Izquierdo-Villalba}, David},
        title = "{The rise and fall of bars in disc galaxies from z = 1 to z = 0. The role of environment}",
      journal = {\aap},
         year = 2024,
        month = apr,
       volume = {684},
          eid = {A179},
        pages = {A179},
          doi = {10.1051/0004-6361/202349003},
archivePrefix = {arXiv},
       eprint = {2401.15215},
 primaryClass = {astro-ph.GA},
       adsurl = {https://ui.adsabs.harvard.edu/abs/2024A&A...684A.179R}
}

@ARTICLE{2019MNRAS.483.2721P,
       author = {{Peschken}, Nicolas and {{\L}okas}, Ewa L.},
        title = "{Tidally induced bars in Illustris galaxies}",
      journal = {\mnras},
         year = 2019,
        month = feb,
       volume = {483},
       number = {2},
        pages = {2721-2735},
          doi = {10.1093/mnras/sty3277},
archivePrefix = {arXiv},
       eprint = {1804.06241},
 primaryClass = {astro-ph.GA},
       adsurl = {https://ui.adsabs.harvard.edu/abs/2019MNRAS.483.2721P}
}

@ARTICLE{2022MNRAS.514.1006I,
       author = {{Izquierdo-Villalba}, David and {Bonoli}, Silvia and {Rosas-Guevara}, Yetli and {Springel}, Volker and {White}, Simon D.~M. and {Zana}, Tommaso and {Dotti}, Massimo and {Spinoso}, Daniele and {Bonetti}, Matteo and {Lupi}, Alessandro},
        title = "{Disc instability and bar formation: view from the IllustrisTNG simulations}",
      journal = {\mnras},
         year = 2022,
        month = jul,
       volume = {514},
       number = {1},
        pages = {1006-1020},
          doi = {10.1093/mnras/stac1413},
archivePrefix = {arXiv},
       eprint = {2203.07734},
 primaryClass = {astro-ph.GA},
       adsurl = {https://ui.adsabs.harvard.edu/abs/2022MNRAS.514.1006I}
}

@ARTICLE{2025ApJ...978...37A,
       author = {{Ansar}, Sioree and {Pearson}, Sarah and {Sanderson}, Robyn E. and {Arora}, Arpit and {Hopkins}, Philip F. and {Wetzel}, Andrew and {Cunningham}, Emily C. and {Quinn}, Jamie},
        title = "{Bar Formation and Destruction in the FIRE-2 Simulations}",
      journal = {\apj},
         year = 2025,
        month = jan,
       volume = {978},
       number = {1},
          eid = {37},
        pages = {37},
          doi = {10.3847/1538-4357/ad8b45},
archivePrefix = {arXiv},
       eprint = {2309.16811},
 primaryClass = {astro-ph.GA},
       adsurl = {https://ui.adsabs.harvard.edu/abs/2025ApJ...978...37A}
}

@ARTICLE{2024MNRAS.529..979L,
       author = {{L{\'o}pez}, Paula D. and {Scannapieco}, Cecilia and {Cora}, Sof{\'\i}a A. and {Gargiulo}, Ignacio D.},
        title = "{Unveiling the origins of galactic bars: insights from barred and unbarred galaxies}",
      journal = {\mnras},
         year = 2024,
        month = apr,
       volume = {529},
       number = {2},
        pages = {979-998},
          doi = {10.1093/mnras/stae576},
archivePrefix = {arXiv},
       eprint = {2403.13061},
 primaryClass = {astro-ph.GA},
       adsurl = {https://ui.adsabs.harvard.edu/abs/2024MNRAS.529..979L}
}

@ARTICLE{2025MNRAS.538.1587F,
       author = {{Fragkoudi}, Francesca and {Grand}, Robert J.~J. and {Pakmor}, R{\"u}diger and {G{\'o}mez}, Facundo and {Marinacci}, Federico and {Springel}, Volker},
        title = "{Bar formation and evolution in the cosmological context: inputs from the Auriga simulations}",
      journal = {\mnras},
         year = 2025,
        month = apr,
       volume = {538},
       number = {3},
        pages = {1587-1608},
          doi = {10.1093/mnras/staf389},
archivePrefix = {arXiv},
       eprint = {2406.09453},
 primaryClass = {astro-ph.GA},
       adsurl = {https://ui.adsabs.harvard.edu/abs/2025MNRAS.538.1587F}
}

@ARTICLE{2001JPhA...3410029S,
       author = {{Skokos}, Ch},
        title = "{Alignment indices: a new, simple method for determining the ordered or chaotic nature of orbits}",
      journal = {Journal of Physics A Mathematical General},
         year = 2001,
        month = nov,
       volume = {34},
       number = {47},
        pages = {10029-10043},
          doi = {10.1088/0305-4470/34/47/309},
       adsurl = {https://ui.adsabs.harvard.edu/abs/2001JPhA...3410029S}
}

@ARTICLE{2014A&C.....5...19C,
       author = {{Carpintero}, D.~D. and {Maffione}, N. and {Darriba}, L.},
        title = "{LP-VIcode: A program to compute a suite of variational chaos indicators}",
      journal = {Astronomy and Computing},
         year = 2014,
        month = jul,
       volume = {5},
        pages = {19-27},
          doi = {10.1016/j.ascom.2014.04.001},
archivePrefix = {arXiv},
       eprint = {1404.2152},
 primaryClass = {nlin.CD},
       adsurl = {https://ui.adsabs.harvard.edu/abs/2014A&C.....5...19C}
}




\appendix

\section{Chaos and spectral entropy}
\label{sec:Apen_A}

The chaotic nature of orbits in galactic dynamics has been widely studied over the years. There are several ways to characterize or measure this chaos. For example, the Poincaré surface can be used to describe it qualitatively, while methods like the maximal Lyapunov exponent \citep{2014A&C.....5...19C} or GALI$_2$ \citep{2001JPhA...3410029S} provide quantitative measures.

Another way to assess the chaoticity of an orbit is through its Fourier spectrum. As indicated in \citet{1998MNRAS.298....1C} and references therein,
the time series of a regular orbit’s coordinates produces a Fourier spectrum made up of discrete peaks. The frequencies of these peaks are linear combinations of a few base frequencies, which correspond to the orbit’s motion in angle variables. If the orbit is closed, there will be only one base frequency. In contrast, a chaotic orbit results in a continuous spectrum without well-defined discrete peaks.

By measuring how discrete or continuous the spectrum is for each coordinate of an orbit, we can estimate how chaotic the orbit is. To do this, we chose to calculate the Shannon entropy \citep{information_theory} using the positive-frequency part of the amplitude spectrum for each Cartesian coordinate. We focused only on the positive frequencies because the amplitude of the Fourier spectrum is symmetric. The Shannon entropy in the $i$-th coordinate is calculated as:

\begin{equation}
    S_i(\omega)=-\sum_{\omega \in \mathcal{X}} p_i(\omega) \log_2 p_i(\omega),
\end{equation}

\noindent where $\mathcal{X}$ denotes the set of all positive frequencies obtained from the Fourier transform. The probability distribution $p_i(\omega)$ is defined as the normalized amplitude spectrum:

\begin{equation}
    p_i(\omega) = \frac{|F_i(\omega)|}{\sum_{\omega \in \mathcal{X}} |F_{i}(\omega)|},
\end{equation}

\noindent where $F_i(\omega)$ is Fourier transform for the $i$-th coordinate. To illustrate the application of the spectral entropy method, we present two representative cases: one regular and one chaotic orbit.

\begin{figure}
  \includegraphics[width=\columnwidth]{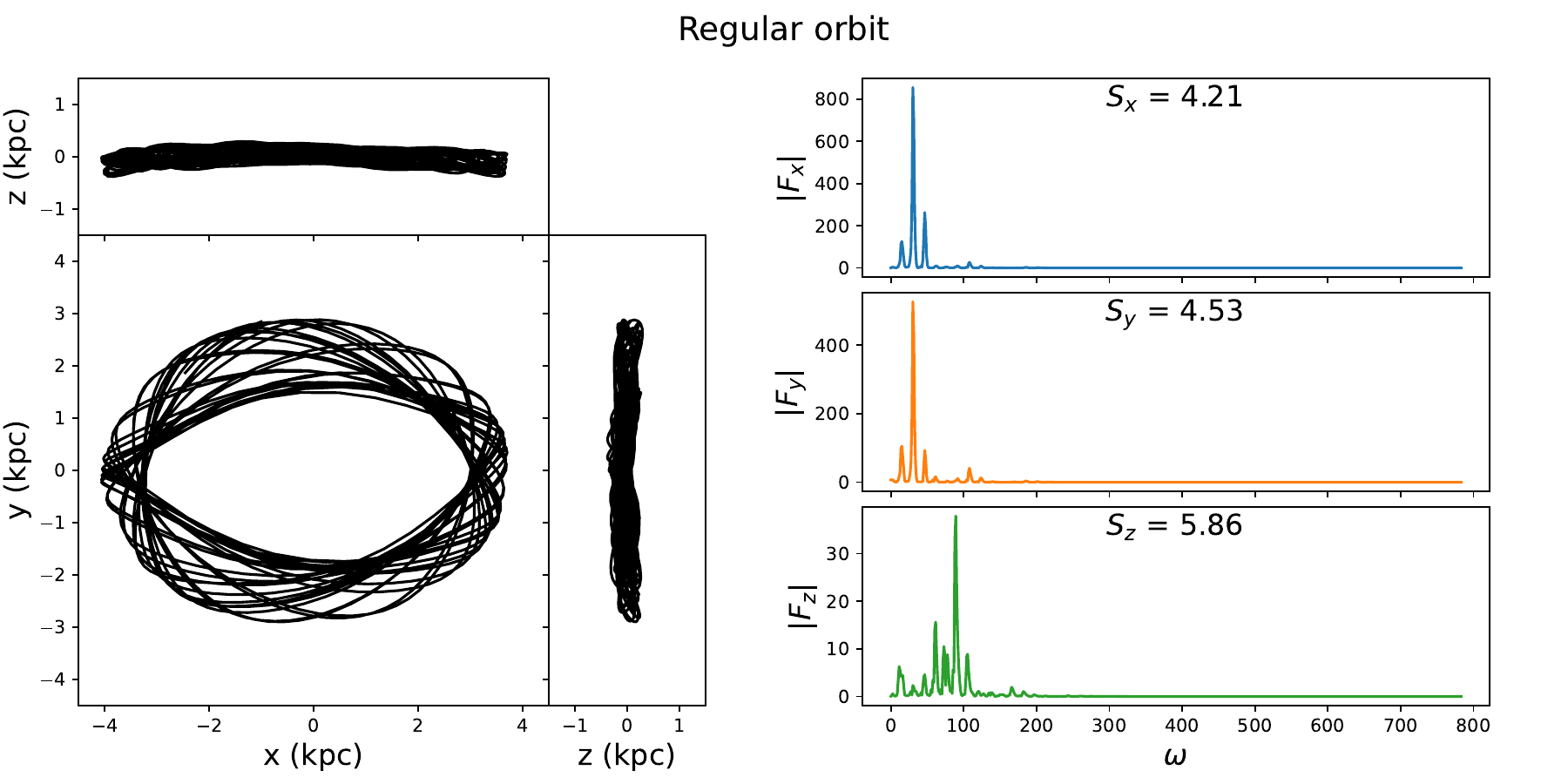}
  \caption{Example of a regular orbit from the AGAMA simulation of $A\lambda03$ at $t = 2.94$ Gyr. The left panels show the projections of the orbit in the $x-y$, $x-z$, and $z-y$ planes. The right panel displays the corresponding Fourier amplitude spectra $|F_i(\omega)|$ for each Cartesian coordinate. The Shannon entropies are also shown above the right panel.}
  \label{fig:reg_orb}
\end{figure}

\begin{figure}
  \includegraphics[width=\columnwidth]{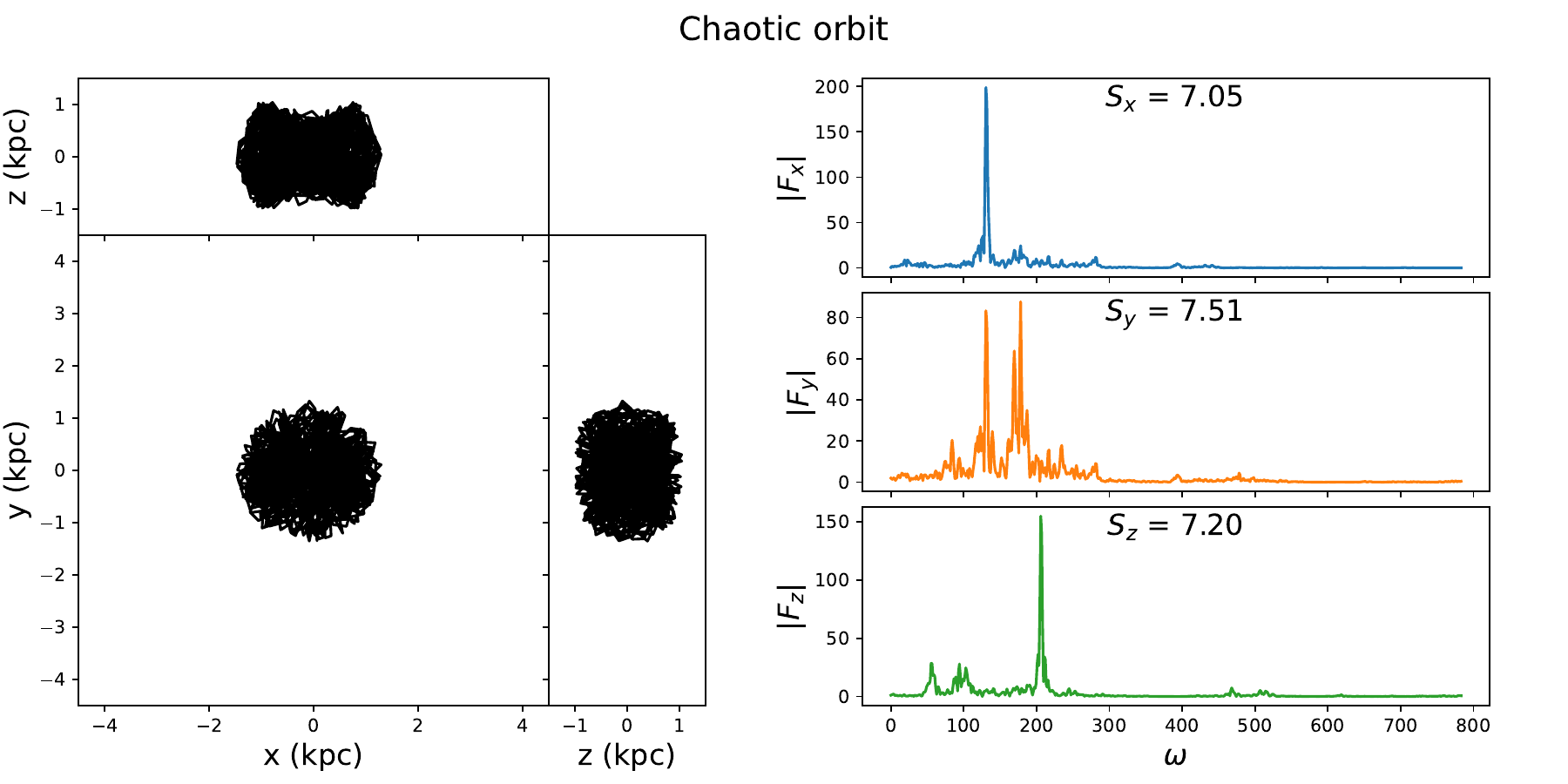}
  \caption{Same example as Fig. \ref{fig:reg_orb} but for a chaotic orbit.}
  \label{fig:cao_orb}
\end{figure}

Figures~\ref{fig:reg_orb} and \ref{fig:cao_orb} present two representative orbits extracted from the AGAMA simulation of $A\lambda03$ at $t=2.94$ Gyr, together with the Fourier amplitude spectra of their Cartesian coordinates and the corresponding Shannon entropies. One orbit is regular, while the other is chaotic. As shown in these figures, chaotic orbits exhibit systematically higher spectral entropy values in all coordinates compared to regular ones. This trend reflects the broader and more complex frequency content of chaotic motion. The clear difference in entropy between the two cases supports the use of spectral entropy as a quantitative indicator for distinguishing regular from chaotic dynamics.

To ensure a robust classification, we define the minimal spectral entropy,

\begin{equation}
    S_{\min} = \min{\big(S_x, S_y, S_z\big)},
\end{equation}

\noindent and adopt it as our diagnostic quantity. By doing so, chaos is identified only when it affects the orbit globally, rather than arising from localized behaviour or numerical artifacts in a single coordinate.

It is worth noting that the value of the calculated entropy is sensitive to some parameters: the total integration time (5 Gyr in our case, for the frozen potentials), the time resolution of the orbit sampling (4 Myr), and the choice of window function applied to the time series prior to the Fourier transform (we use a Blackman window). These factors influence the shape and amplitude of the resulting Fourier spectrum and, consequently, the entropy values derived from it. However, this sensitivity does not undermine the use of spectral entropy as a reliable tool to estimate the chaoticity of an orbit, as the relative distinction between regular and chaotic behaviour remains robust across different configurations.

We also verified that the calculated entropy values are not significantly affected by the numerical spatial resolution of the simulation. Since the spectral entropy is derived from orbit integrations in frozen potentials, its value is primarily controlled by the temporal sampling and integration parameters discussed above.

\section{Calculating the \texorpdfstring{$\mathbf{x_1}$}{TEXT} threshold}
\label{sec:Apen_B}

\begin{figure}
  \includegraphics[width=\columnwidth]{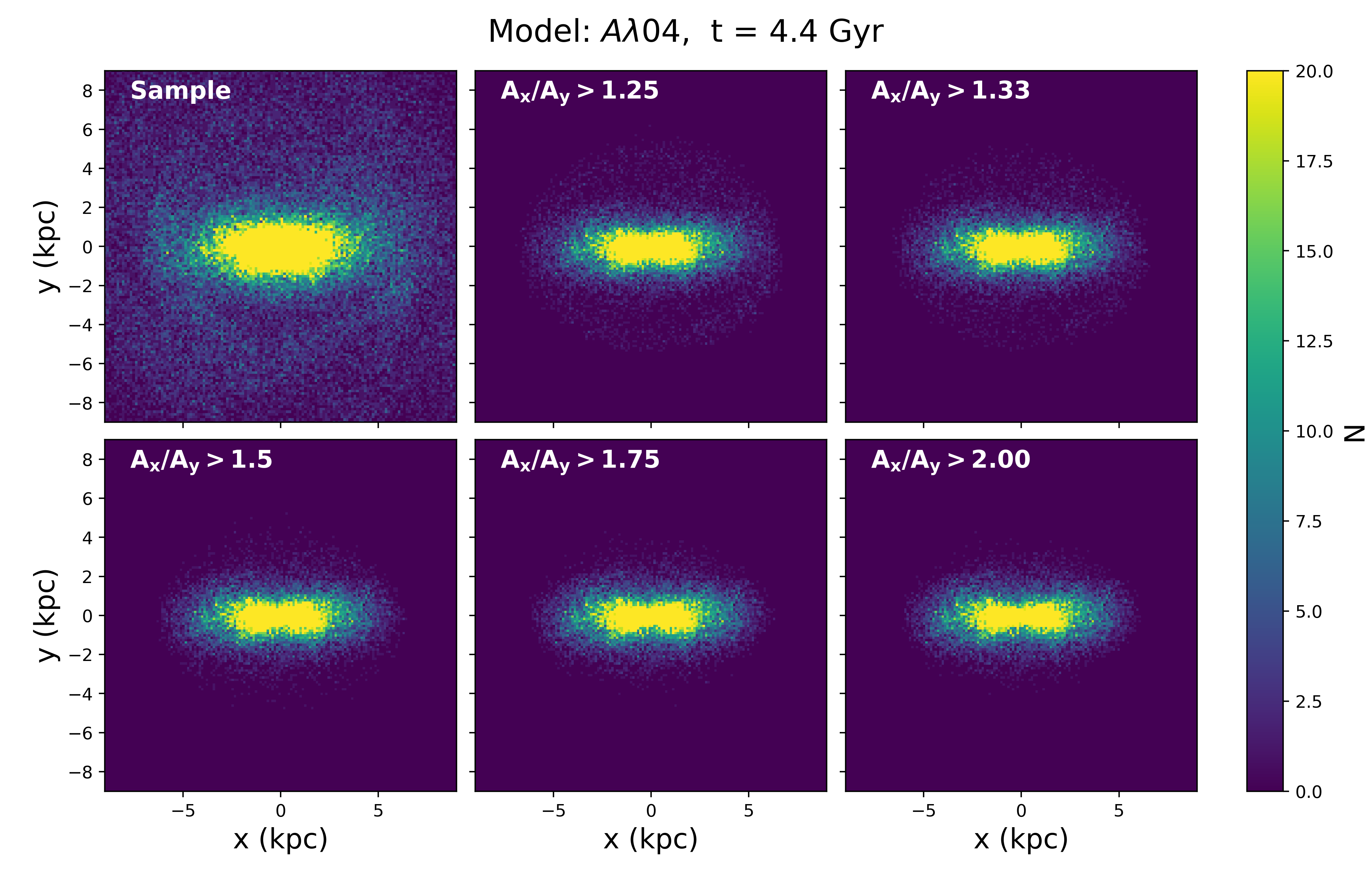}
  \caption{Face-on 2D histograms of a 100,000-particle sample from model $A\lambda04$ at $t = 4.4$ Gyr. The upper-left panel shows the full sample, while the remaining panels display subsets of particles belonging to the $x_1$ family, selected by increasing axis ratio thresholds ($A_x/A_y$).}
  \label{fig:2D_ra_L40}
\end{figure}

As discussed in Sec. \ref{sec:frec_an}, the “$x_1$ range” of the distribution in Fig. \ref{fig:hist_amp} does not exhibit a sharp or consistent minimum. This makes it difficult to define a clear threshold for identifying $x_1$ orbits directly from that distribution. To address this, we examined the two-dimensional spatial distributions of elliptical-like particles under different $A_x/A_y$ thresholds, across both models and multiple snapshots. Figure \ref{fig:2D_ra_L40} shows the case of model $A\lambda04$ at $t = 4.4$ Gyr, which clearly illustrates our approach.

\begin{figure}
\centering
  \includegraphics[width=0.6\columnwidth]{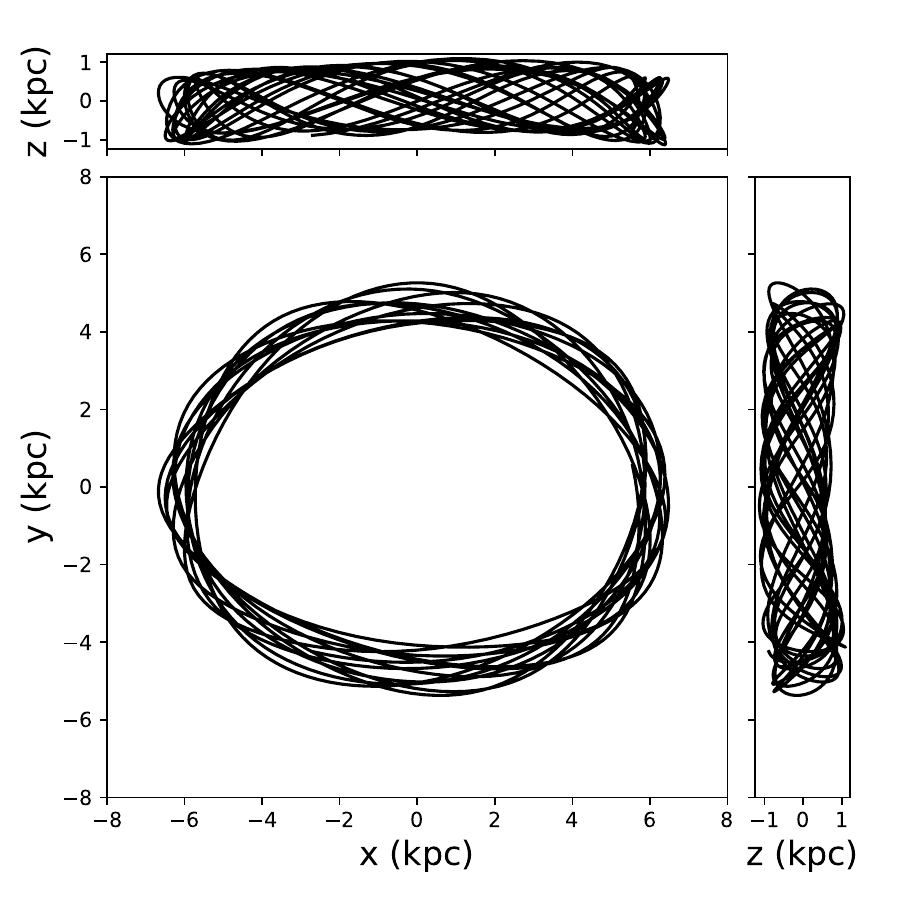}
  \caption{Example of an orbit belonging to the surrounding “ring” structure identified in Fig. \ref{fig:2D_ra_L40}. The panels show the projections onto the $(x,y)$, $(x,z)$, and $(y,z)$ planes. This type of orbit circulates around the $x_1$ family but does not contribute to the bar structure, and was therefore excluded from the $x_1$ classification.}
  \label{fig:rara_orb}
\end{figure}

As shown in Fig. \ref{fig:2D_ra_L40}, the distribution in the upper-middle panel highlights the $x_1$ family of particles. However, it also reveals a surrounding ring$-$like structure. By inspecting individual orbits within this region, we found that the ring corresponds to particles orbiting around the $x_1$ structure rather than belonging to it. An example of such an orbit is presented in Fig. \ref{fig:rara_orb}. Since these orbits cannot be considered part of the $x_1$ family or the bar itself, we excluded them from our analysis. To achieve this, we adopted a threshold of $A_x/A_y = 1.5$ (distribution shown in the lower-left panel in Fig. \ref{fig:2D_ra_L40}), which effectively removes these spurious orbits while retaining the majority of genuine $x_1$ orbits. This criterion provides a consistent and reliable definition of the $x_1$ family, which we adopt throughout the rest of the analysis.

\section{Spatial and temporal resolution tests.}
\label{sec:Apen_C}

As described in the AGAMA documentation \citep{2018arXiv180208255V}, the {\tt Multipole} potential expansion represents the gravitational potential as a sum of spherical-harmonic functions in the angular coordinates, multiplied by arbitrary radial functions. This approach is well suited for representing approximately spherical or mildly triaxial mass distributions.

In contrast, the {\tt CylSpline} potential represents the gravitational potential as a Fourier expansion in the azimuthal angle, with spline interpolation in the meridional plane. It is therefore particularly appropriate for flattened systems such as stellar discs.

In $N$-body simulations, these two expansions naturally approximate different galactic components: the {\tt Multipole} expansion is typically used for the dark matter halo, while the {\tt CylSpline} expansion provides an efficient representation of the stellar disc.

Both expansions rely on a finite number of grid nodes that define the radial (or cylindrical) resolution of the potential. In the case of {\tt Multipole}, the grid nodes are distributed in spherical radius. Whereas for {\tt CylSpline}, the grid is defined in cylindrical coordinates.

For the fiducial model used in this work, we adopted $N_s = 25$ grid nodes for both potential expansions in order to maintain a comparable numerical resolution. To assess the sensitivity of our results to this choice, we performed additional tests varying the number of grid nodes. Specifically, we repeated the calculations for $N_s = 50$ and $N_s = 100$. We then examined the distribution of the frequency ratio $\omega_R / \omega_x$ to determine whether increasing the number of nodes produces any significant change in its structure.

In addition to the spatial resolution of the potential, the temporal sampling of the orbital integration in the frozen potential also affects the accuracy of the frequency determination. Since the fundamental frequencies are calculated via Fourier transforms of the orbital coordinates, the number of temporal sampling points directly impacts the spectral resolution. To assess this effect, we performed an analogous test by varying the number of time samples used in the Fourier analysis. Starting from the fiducial value $N_t = 1{,}250$, we increased the sampling to $N_t = 2{,}500$ and $N_t = 5{,}000$ points over the same total integration time.

In order to isolate the effect of temporal resolution alone, all tests were conducted using the same total integration time (5 Gyr) and the same randomly selected subset of 100{,}000 particles.

\begin{figure}
\centering
  \includegraphics[width=0.9\columnwidth]{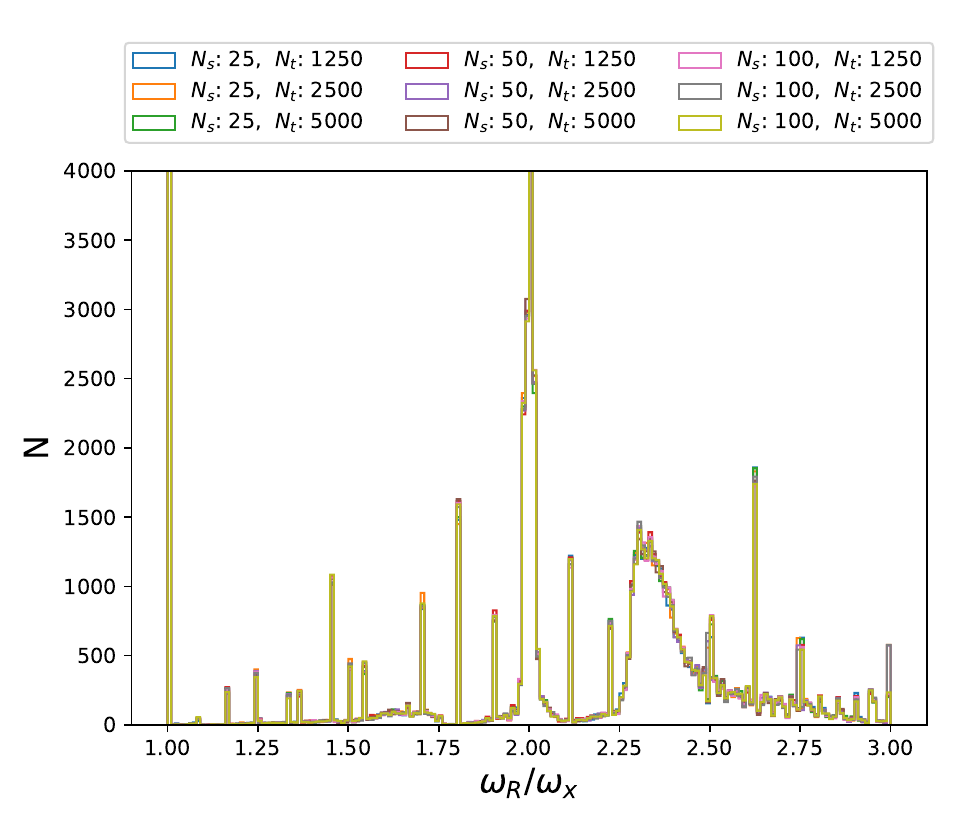}
  \caption{Distribution of the frequency ratio $\omega_R/\omega_x$ for model $A\lambda03$, evaluated at snapshot 3,000, for different spatial resolutions ($N_s$) and temporal sampling values ($N_t$). The comparison includes $N_s = 25, 50, 100$ and $N_t = 1{,}250, 2{,}500, 5{,}000$.}
  \label{fig:test_res}
\end{figure}

Figure~\ref{fig:test_res} presents the distribution of the frequency ratio $\omega_R/\omega_x$ in the range $1 \leq \omega_R/\omega_x \leq 3$ for model $A\lambda03$, evaluated at snapshot 3,000, for different combinations of spatial ($N_s$) and temporal ($N_t$) resolutions. The overall structure of the distribution (including the location, width, and relative prominence of the main peaks) remains unchanged as either the number of grid nodes or the number of temporal sampling points is increased.

Only negligible variations at the level of statistical noise are observed. This indicates that the fiducial choice ($N_s = 25$, $N_t = 1{,}250$) is sufficient to ensure numerical convergence with respect to both the spatial representation of the potential and the temporal resolution used in the Fourier frequency analysis.

On the other hand, increasing $N_s$ and $N_t$ leads to a increase in computational cost without producing statistically significant changes in the frequency distribution. We therefore adopt the fiducial values as an optimal compromise between accuracy and efficiency.

In summary, the stability of the $\omega_R/\omega_x$ distribution under variations in both spatial grid resolution and temporal sampling demonstrates that our results are numerically converged. The fiducial configuration ($N_s = 25$, $N_t = 1{,}250$) provides a reliable representation of the potential and sufficient spectral resolution for frequency determination, while avoiding unnecessary computational overhead. All results presented in the main text are therefore based on these fiducial values.


\bsp	
\label{lastpage}
\end{document}